\documentclass{elsarticle}

\usepackage[reviewcopy]{adndt}
\usepackage{longtable}

\usepackage{amsmath}

\usepackage{amssymb}

\biboptions{square,sort&compress}
\bibpunct[]{[}{]}{,}{n}{}{;}
\begin{document}

\begin{frontmatter}

\journal{Atomic Data and Nuclear Data Tables}


\title{Discovery of Yttrium, Zirconium, Niobium, Technetium, and Ruthenium Isotopes}

\author{A. Nystrom}
\author{M. Thoennessen\corref{cor1}}\ead{thoennessen@nscl.msu.edu}

 \cortext[cor1]{Corresponding author.}

 \address{National Superconducting Cyclotron Laboratory and \\ Department of Physics and Astronomy, Michigan State University, \\ East Lansing, MI 48824, USA}

\begin{abstract}
Currently, thirty-four yttrium, thirty-five zirconium, thirty-four niobium, thirty-five technetium, and thirty-eight ruthenium isotopes have been observed and the discovery of these isotopes is discussed here. For each isotope a brief synopsis of the first refereed publication, including the production and identification method, is presented.
\end{abstract}

\end{frontmatter}





\newpage
\tableofcontents
\listofDtables

\vskip5pc

\section{Introduction}\label{s:intro}

The discovery of yttrium, zirconium, niobium, technetium, and ruthenium isotopes is discussed as part of the series summarizing the discovery of isotopes, beginning with the cerium isotopes in 2009 \cite{2009Gin01}. Guidelines for assigning credit for discovery are (1) clear identification, either through decay-curves and relationships to other known isotopes, particle or $\gamma$-ray spectra, or unique mass and Z-identification, and (2) publication of the discovery in a refereed journal. The authors and year of the first publication, the laboratory where the isotopes were produced as well as the production and identification methods are discussed. When appropriate, references to conference proceedings, internal reports, and theses are included. When a discovery includes a half-life measurement the measured value is compared to the currently adopted value taken from the NUBASE evaluation \cite{2003Aud01} which is based on the ENSDF database \cite{2008ENS01}. In cases where the reported half-life differed significantly from the adopted half-life (up to approximately a factor of two), we searched the subsequent literature for indications that the measurement was erroneous. If that was not the case we credited the authors with the discovery in spite of the inaccurate half-life. All reported half-lives inconsistent with the presently adopted half-life for the ground state were compared to isomers half-lives and accepted as discoveries if appropriate following the criterium described above.

The first criterium excludes measurements of half-lives of a given element without mass identification. This affects mostly isotopes first observed in fission where decay curves of chemically separated elements were measured without the capability to determine their mass. Also the four-parameter measurements (see, for example, Ref. \cite{1970Joh01}) were, in general, not considered because the mass identification was only $\pm$1 mass unit.

The second criterium affects especially the isotopes studied within the Manhattan Project. Although an overview of the results was published in 1946 \cite{1946TPP01}, most of the papers were only published in the Plutonium Project Records of the Manhattan Project Technical Series, Vol. 9A, ''Radiochemistry and the Fission Products,'' in three books by Wiley in 1951 \cite{1951Cor01}. We considered this first unclassified publication to be equivalent to a refereed paper.
Good examples why publications in conference proceedings should not be considered are $^{118}$Tc and $^{120}$Ru which had been reported as being discovered in a conference proceeding \cite{1996Cza01} but not in the subsequent refereed publication \cite{1997Ber01}.

The initial literature search was performed using the databases ENSDF \cite{2008ENS01} and NSR \cite{2008NSR01} of the National Nuclear Data Center at Brookhaven National Laboratory. These databases are complete and reliable back to the early 1960's. For earlier references, several editions of the Table of Isotopes were used \cite{1940Liv01,1944Sea01,1948Sea01,1953Hol02,1958Str01,1967Led01}. A good reference for the discovery of the stable isotopes was the second edition of Aston's book ``Mass Spectra and Isotopes'' \cite{1942Ast01}.

\section{Discovery of $^{76-109}$Y}

Thirty-four yttrium isotopes from A = 76--109 have been discovered so far; these include 1 stable, 13 neutron-deficient and 20 neutron-rich isotopes. According to the HFB-14 model \cite{2007Gor01}, $^{120}$Y should be the last odd-odd particle stable neutron-rich nucleus while the odd-even particle stable neutron-rich nuclei should continue through $^{131}$Y. The proton dripline has most likely been reached at $^{76}$Y, however, $^{75}$Y and $^{74}$Y could still have half-lives longer than 10$^{-9}$~ns \cite{2004Tho01}. Thus, about 19 isotopes have yet to be discovered corresponding to 40\% of all possible yttrium isotopes.

\begin{figure}
	\centering
	\includegraphics[scale=.5]{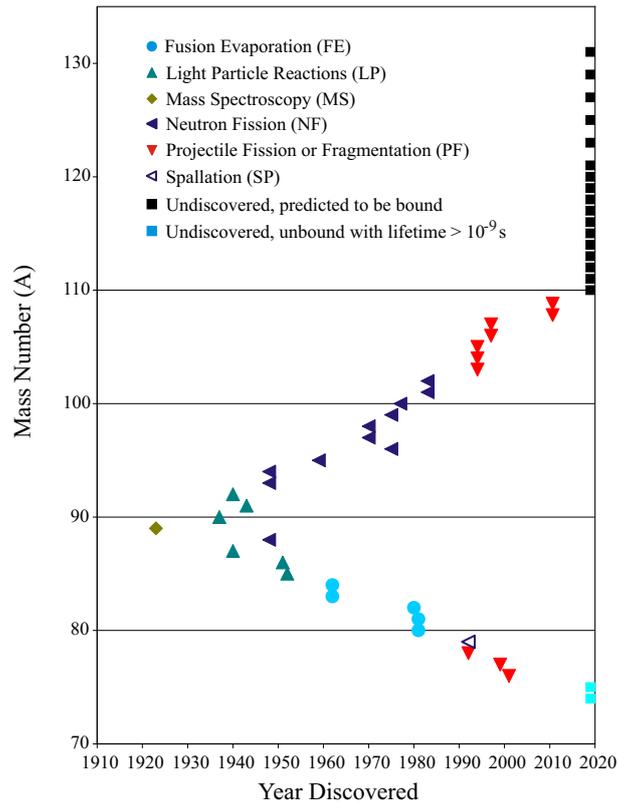}
	\caption{Yttrium isotopes as a function of time when they were discovered. The different production methods are indicated. The solid black squares on the right hand side of the plot are isotopes predicted to be bound by the HFB-14 model. On the proton-rich side the light blue squares correspond to unbound isotopes predicted to have lifetimes larger than $\sim 10^{-9}$~s.}
\label{f:year-y}
\end{figure}

Figure \ref{f:year-y} summarizes the year of first discovery for all yttrium isotopes identified by the method of discovery. The range of isotopes predicted to exist is indicated on the right side of the figure. The radioactive yttrium isotopes were produced using fusion evaporation reactions (FE), light-particle reactions (LP), neutron induced fission (NF), spallation (SP), and projectile fragmentation or fission (PF). The stable isotope was identified using mass spectroscopy (MS). Light particles also include neutrons produced by accelerators. The discovery of each yttrium isotope is discussed in detail and a summary is presented in Table 1.

\subsection{$^{76}$Y}\vspace{0.0cm}

In the 2001 article ``Synthesis and Halflives of Heavy Nuclei Relevant for the rp-Process,'' Kienle et al. first reported the existence of $^{76}$Y \cite{2001Kie01}. A 1~A$\cdot$GeV $^{112}$Sn beam from the synchrotron SIS at GSI, Germany, bombarded a beryllium target. $^{76}$Y was identified with the FRS fragment separator by both magnetic deflection and by measuring the energy loss and time-of-flight of the fragments. ``The spectra show the previously unobserved T = $-$1 nuclei $^{76}$Y (2 events) and $^{78}$Zr (one event) and demonstrate the absence of $^{81}$Nb...''

\subsection{$^{77}$Y}\vspace{0.0cm}

$^{77}$Y was first reported in ``Observation of the Z = N + 1 Nuclei $_{39}^{77}$Y, $_{40}^{79}$Zr, and $_{42}^{83}$Mo'' in 1999 by Janas et al. \cite{1999Jan01}. At GANIL, France, nickel targets were bombarded with a 60 MeV/nucleon $^{92}$Mo beam. $^{77}$Y was separated with the LISE3 spectrometer and the kinetic energy, energy loss, and time-of-flight were measured. ``In the region expected for $^{77}_{39}$Y, [the figure] clearly shows a peak indicating that the half-life of this isotope is longer than 0.5~ms, the flight time through the LISE3 spectrometer.''

\subsection{$^{78}$Y}\vspace{0.0cm}

The discovery of $^{78}$Y is credited to Yennello et al. with their 1992 paper ``New Nuclei Along the Proton-Drip Line Near Z = 40,'' \cite{1992Yen01}.  At the National Superconducting Cyclotron Laboratory at Michigan State University, a 70~MeV/A $^{92}$Mo beam was produced by the K1200 cyclotron and impinged on a $^{58}$Ni target. $^{78}$Y was identified with the A1200 fragment analyzer by measuring the time-of-flight and energy loss of the fragments. ``The mass spectra for residues with Z from 39 to 44 are shown in [the figure] with the new isotopes marked by arrows. Although both $^{84}$Mo and $^{86}$Mo have been previously observed, no reference to the identification of $^{85}$Mo was found. The other new isotopes observed in this study are $^{78}$Y, $^{82}$Nb, $^{86}$Tc, and $^{89,90}$Ru.''

\subsection{$^{79}$Y}\vspace{0.0cm}

Grawe et al. discovered $^{79}$Y in ``Study of the $\beta$$^{+}$/EC Decay of the Neutron Deficient Nuclei $^{76,78}$Sr and $^{79}$Y,'' in 1992 \cite{1992Gra01}. $^{79}$Y was produced in spallation reactions on a niobium foil at CERN, Switzerland. $^{79}$Y was identified with the ISOLDE mass separator and $\gamma$-ray spectra were measured with a HPGe detector. ``The weighted averages of half lives determined for various $\gamma$-rays in the daughter nuclei are t$_{1/2}$ = 8.9(3)~s, 159(8)~s and 14.4(15)~s for $^{76,78}$ Sr and $^{79}$Y, respectively.'' This half-life for $^{79}$Y is included in the accepted average value of 14.8(6)~s. The observation of $^{79}$Y had previously been reported in a conference proceeding \cite{1987Lob01} and Mukai et al. submitted their independent results only two months later \cite{1992Muk01}.

\subsection{$^{80,81}$Y}\vspace{0.0cm}

The first observations of $^{80}$Y and $^{81}$Y were reported in ``New Isotope $^{80}$Y, and the Decays of $^{79}$Sr, $^{81}$Y, and $^{82}$Y'' by Lister et al. in 1981 \cite{1981Lis01}. The Brookhaven Tandem Van de Graaff accelerator provided beams of 75-110 MeV $^{24}$Mg and 91-110 MeV $^{25}$Mg which then bombarded enriched $^{58}$Ni targets. $^{80}$Y and $^{81}$Y were produced in the fusion evaporation reactions $^{58}$Ni($^{24}$Mg,\textit{pn})$^{80}$Y and $^{58}$Ni($^{25}$Mg,\textit{pn})$^{81}$Y, respectively. The recoil products were thermalized and deposited onto a Mylar Tape loop, where spectroscopic measurements were made. ``The present work has resulted in the identification of a new neutron-deficient nuclide, $^{80}$Y. Extensive decay scheme information has been obtained for $^{80}$Y and the previously poorly characterized radioactivities of $^{79}$Sr, $^{81}$Y, and $^{82}$Y.'' The measured half-life of 33.8$\pm$0.6~s for $^{80}$Y is consistent with the presently accepted value of 30.1(5)~s and the half-life of 72.0$\pm$1.5~s for $^{81}$Y is included in the calculation of the current accepted value of 70.4(10)~s. Previously, a 5~min half-life included in the 1978 edition of the Table of Isotopes \cite{1978Led01} was only reported in an annual report \cite{1970Has01}. This half-life was evidently incorrect.

\subsection{$^{82}$Y}\vspace{0.0cm}

In ``Mass Excess Values of $^{79}$Sr, $^{82}$Y and $^{83}$Y obtained from Q$_{\beta}$ measurements,'' Deprun et al. described the observation of $^{82}$Y in 1980 \cite{1980Dep01}. A $^{32}$S beam with energies ranging from 100 to 160~MeV on $^{54}$Fe and $^{58}$Ni targets produced $^{82}$Y in fusion-evaporation reactions. The reaction products were collected with a helium jet system and the masses were identified by time-of-flight. ``By assuming that there are $\beta^+$ transitions from [the] ground state of $^{82}$Y to the ground state of $^{82}$Sr, the endpoint energy deduced from the $\beta^+$ spectra is 6.60$\pm$0.1~MeV and $\Delta $M=7.62$\pm$0.1~MeV.'' The measured half-life of 9.5(5)~s is close to the accepted value of 8.30(20)~s. Much longer half-lives of 70(10)~min \cite{1952Car01}, 9(3)~min \cite{1962Max01}, 12.3(2)~min \cite{1963But01}, 7.4(1)~min \cite{1963Dos01}, and 7.5(1)~min \cite{1965Nie01} were evidently incorrect.

\subsection{$^{83,84}$Y}\vspace{0.0cm}

$^{83}$Y and $^{84}$Y were discovered by Maxia et al. in the 1962 paper ``The Neutron-Deficient Yttrium Isotopes $^{82}$Y, $^{83}$Y and $^{84}$Y'' \cite{1962Max01}. A beam of approximately 120~MeV $^{12}$C from the Berkeley Heavy-Ion Linear Accelerator was used to bombard powdered arsenic metal targets. The isotopes were produced in the fusion evaporation reactions $^{75}$As($^{12}$C,4n)$^{83}$Y and $^{75}$As($^{12}$C,3n)$^{84}$Y. Half-lives were determined by measuring $\beta$-particles with an end-window, flowing methane-proportional counter. ``Consideration of the data from several experiments utilizing different production modes of yttrium and different milk times leads to an average value of 8$\pm$2 min for the half-life of $^{83}$Y.'' and ``Yttrium-84: When the yttrium fraction obtained from the bombardment of arsenic with C$^{12}$ ions was chemically repurified approximately 2~hr after the irradiation, the gamma-ray spectrum decayed with 39$\pm$2 min.'' The half-life of 8(2)~min for $^{83}$Y agrees with the currently accepted values of 7.08(6)~min and the half-life of 39(2)~min for $^{84}$Y is included in the calculation of the presently accepted average value of 39.5(8)~min. Previously reported half-lives of 3.5(5)~h \cite{1952Car01} for  $^{83}$Y and 3.7(1)~h \cite{1949Rob01} for $^{84}$Y were evidently incorrect.

\subsection{$^{85}$Y}\vspace{0.0cm}

The discovery of $^{85}$Y is described by Carotto and Wiig in their paper ``Three New Neutron Deficient Isotopes of Yttrium'' published in 1952 \cite{1952Car01}. A target of yttrium oxide covered in an aluminum envelope was bombarded with 130 and 240 MeV protons at the Rochester 130-inch cyclotron. Decay curves were recorded with a Geiger-M\"{u}ller tube following chemical separation. ``These experiments give the following values for the half lives of the yttrium isotopes: Y$^{82}$ 70$\pm$10 minutes, Y$^{83}$ 3.5$\pm$0.5 hours, Y$^{85}$ 5$\pm$1 hours.'' The half-life of $^{85}$Y corresponds to an isomeric state.
It should be noted that the half-lives for $^{82}$Y and $^{83}$Y were incorrect, however, the known half-lives of $^{86}$Y, $^{87}$Y, and $^{88}$Y, were reproduced. Subsequent measurements \cite{1962Max01} did not question the results of Carotto and Wiig.

\subsection{$^{86}$Y}\vspace{0.0cm}

Hyde and O'Kelley first identified $^{86}$Y in their paper, ``Radiochemical and Spectrometer Studies of Several Neutron-Deficient Zirconium Isotopes and Their Decay Products,'' in 1951 \cite{1951Hyd01}. Thin strips of niobium metal were bombarded with 100 MeV protons from the Berkeley 184-inch synchrocyclotron. The spallation products were separated through rapid radiochemistry and studied with a beta-ray spectrometer. ``Zr$^{86}$ is a 17$\pm$2-hour orbital electron capturing isotope decaying into Y$^{86}$, which in turn disintegrates into stable Sr$^{86}$ with a half-life of 14.6$\pm$0.2 hours by the emission of positrons.'' This half-life agrees with the presently accepted value of 14.74(2)~h. A 105-day activity had previously been incorrectly assigned to $^{86}$Y \cite{1940DuB01}. Subsequently this half-life (100~days) had been reported by other authors without questioning the mass assignment \cite{1940Pec01,1941Dow01,1941Reg01}.

\subsection{$^{87}$Y}\vspace{0.0cm}

The first observation of $^{87}$Y was documented by DuBridge and Marshall in their 1940 paper entitled ``Radioactive Isotopes of Sr, Y and Zr'' \cite{1940DuB01}. Strontium targets were bombarded with 6.7~MeV protons at the University of Rochester. The activities of the reaction products were measured with a freon-filled ionization chamber, and the $\beta$ and $\gamma$ decays were measured with a magnetic cloud chamber and a $\beta$-ray spectrograph. ``In addition ... we have observed three other periods produced by protons: 14$\pm$2 hours, 80$\pm$3 hours, and 105$\pm$5 days. Since the 14-hr. and the 80-hr. period are also produced by the bombardment of Sr with deuterons they can be assigned only to Y$^{87}$ or Y$^{88}$. Stewart, Lawson and Cork, however, found only the 2-hr. period as a result of the reaction Y$^{89}$(n,2n). Therefore, we must assume that the 14-hr. and 80-hr. period are isomers of Y$^{87}$.'' The 80(3)~h half-life agrees with the presently accepted value of 79.8(3)~h and the 14(2)~h half-life corresponds to an isomeric state. Half-lives of 14(2)~h and 82(4)~h had been observed previously; however, the assignments to either $^{85}$Y or $^{87}$Y was uncertain \cite{1939Ste01}.

\subsection{$^{88}$Y}\vspace{0.0cm}

In ``Mass Spectrographic Mass Assignment of Radioactive Isotopes,'' Hayden provided the first correct identification of $^{88}$Y in 1948 \cite{1948Hay01}. A 10 mC 108-day yttrium sample produced in the Clinton pile was placed in the ion source of a spectrograph. The mass separated ions were deposited on a photographic plate and the activities measured with a Geiger counter. ``Upon development the original plate showed two lines one mass number apart, and the transfer plate showed one line which corresponded to the lower mass line on the original plate. Therefore the higher mass line was the normal Y$^{89}$O$^+$, and the active line one mass unit lighter was Y$^{88}$O$^+$. Thus the mass of the 108-day yttrium is 88.'' In 1937 Stewart et al. assigned a 120(4)~min half-life incorrectly to $^{88}$Y \cite{1937Ste01}. A 105-day activity had been observed previously, but had been assigned to $^{86}$Y \cite{1940DuB01}. Subsequently this half-life (100~days) had been reported by other authors without questioning the mass assignment \cite{1940Pec01,1941Dow01,1941Reg01}.

\subsection{$^{89}$Y}\vspace{0.0cm}

In 1923 Aston reported the discovery of the only stable yttrium isotope, $^{89}$Y in ``Further Determinations of the Constitution of the Elements by the Method of Accelerated Anode Rays'' \cite{1923Ast01}. No details regarding the mass spectroscopic observation of yttrium were given. ``Yttrium behaved surprisingly well, and yielded unequivocal results indicating that it is a simple element of mass number 89.''

\subsection{$^{90}$Y}\vspace{0.0cm}

The identification of $^{90}$Y was first reported by Pool et al. in the 1937 paper ``A Survey of Radioactivity Produced by High Energy Neutron Bombardment'' \cite{1937Poo01}. The bombardment of yttrium with 20 MeV neutrons resulted in the observation of a half-life of 2.4 days which was assigned to $^{90}$Y. The paper presented the results for a large number of elements and ``The assignments of the periods is tentative and is based upon evidence from the sign of the emitted beta-particle, the chemical separations and known periods from other sources.'' This half-life agrees with the half-life of the ground state (64~h). The observation of $^{90}$Y was also reported independently two months later by Stewart et al. \cite{1937Ste01}.

\subsection{$^{91}$Y}\vspace{0.0cm}

$^{91}$Y was first correctly identified by Seelmann-Eggebert in ``\"Uber einige aktive Yttrium-Isotope'' in 1943 \cite{1943See02}. Zirconium was irradiated with fast Li/D neutrons and several yttrium activities were measured following chemical separation. ``Es konnte festgestellt werden, da\ss\ das mit dem 57 Tage Yttrium isomere 50 Minuten-Yttrium der angeregte Zustand ist, und da\ss\ die Muttersubstanz dieser Isomeren, das 10 Stunden-Strontium, auch bei Bestrahlung des Zirkons mit schnellen Neutronen durch n,$\alpha$ Proze\ss\ entsteht. Die Masse dieser Reihe ist daher 91.'' [It could be determined that the 50-min yttrium isomer of the 57-day yttrium corresponds to the excited state and that the 10-hour strontium mother substance of these isomers could also be produced in the n,$\alpha$ process by bombarding zirconium with fast neutrons. The mass of this chain is thus 91.] The half-life of the ground-state agrees with the presently accepted value of 58.51(6)~days. A 57(3)~day half-life had been previously reported without a mass assignment \cite{1940Hah01} and in 1941 it had been demonstrated that the 57~day and 50~min yttrium activities are isomeric, however, a mass assignment was still not possible \cite{1941Got01}.

\subsection{$^{92}$Y}\vspace{0.0cm}

Sagane et al. reported the discovery of $^{92}$Y in their 1940 paper ``Artificial Radioactivity Induced in Zr and Mo'' \cite{1940Sag01}. Fast neutrons produced in the reaction of deuterons on lithium from the Tokyo cyclotron irradiated zirconium targets. No details were given in the paper and the results were summarized in a table. The measured half-life of 3.0(5)~h is in reasonable agreement with the currently  accepted value of 3.54(1)~h. A 3.5~h half-life had been reported previously for yttrium without a mass assignment \cite{1939Lie01}.

\subsection{$^{93,94}$Y}\vspace{0.0cm}

In 1948, Katcoff et al. published the first identifications of $^{93}$Y and $^{94}$Y in ``Ranges in Air and Mass Identification of Plutonium Fission Fragments'' \cite{1948Kat01}. Plutonium targets were irradiated with neutrons in the Los Alamos homogeneous pile and activities and ranges of the fission fragments were measured following chemical separation. ``From the range-mass curve drawn for well-known masses, definite assignments of 92,93, and 132 were given to 3.5-hr. Y, 10-hr. Y, and 77-hr. Te, respectively. Highly probable assignments of 94 and 134 were given to 20-min. Y and 54-min. I, respectively.'' These half-lives agree with the currently accepted values of 10.18(8)~h and 18.7(1)~min for $^{93}$Y and $^{94}$Y, respectively. Hahn and Strassmann had observed half-lives of 20~min and 11.6~h but could only determine that the mass was larger than 91 \cite{1943Hah01,1943Hah02}.

\subsection{$^{95}$Y}\vspace{0.0cm}

The first observation of $^{95}$Y is reported in ``Radiations of $^{93}$Y and $^{94}$Y and Half-Lives of $^{93}$Sr and $^{94}$Sr,'' in 1959 by Knight et al. \cite{1959Kni01}. Neutrons from the Los Alamos ``Water Boiler'' reactor irradiated $^{235}$U; fission fragments were chemically separated and $\beta$-decay curves were recorded. ``Subtraction of the 10.5~hr $^{93}$Y and 3.62~hr $^{92}$Y components gave a strong 20$\pm$1 min component, $^{94}$Y, and at early times a small amount of a ~9.5~min component which was attributed to $^{95}$Y.'' This half-life is consistent with the accepted half-life of 10.3(1)~min. During the Manhattan project a 10~h yttrium activity had initially been assigned to $^{95}$Y but had later been corrected to $^{93}$Y \cite{1951Sel01}.

\subsection{$^{96}$Y}\vspace{0.0cm}

In the 1975 paper ``Gamma-Ray Energies of Short-Lived Rb, Sr, and Y Isotopes Indicating Delayed Neutron Emission to Excited States,'' Gunther et al. correctly identified $^{96}$Y for the first time \cite{1975Gun01}. A $^{235}$U target was irradiated with neutrons in the Munich research reactor and $^{96}$Y was identified with an on-line mass separator and by $\beta$-$\gamma$ coincidence measurements. `Gamma energies in the level schemes of $^{95,96,97}$Sr, $^{95,96,97}$Y, and $^{95,97}$Zr are reported.'' A previously reported half-life of 2.3(1)~min \cite{1961Val01} was evidently incorrect.

\subsection{$^{97}$Y}\vspace{0.0cm}

In 1970, Eidens et al. described the first observation of $^{97}$Y in ``On-Line Separation and Identification of Several Short-Lived Fission Products: Decay of $^{84}$Se, $^{91}$Kr, $^{97}$Y, $^{99}$Nb, $^{99}$Zr, $^{100,101}$Nb and $^{101}$Zr'' \cite{1970Eid01}. Neutrons from the J\"ulich FRJ-2 reactor irradiated a $^{235}$U target and the fission fragments were identified with a gas-filled on-line mass separator. Beta-$\gamma$- and $\gamma$-$\gamma$-coincidences were recorded. ``Two $\gamma$-lines with energies of 125$\pm$3 keV and 810$\pm$3 keV were found to be coincident in the $\gamma$-$\gamma$ coincidence studies. They were assigned to $^{97}$Y.'' The measured half-life of 1.11(3)~s corresponds to an isomeric state.

\subsection{$^{98}$Y}\vspace{0.0cm}

Gr\"uter et al. observed $^{98}$Y as reported in the 1970 paper ``Identification of $\mu$s-Isomers among Primary Fission Products,'' \cite{1970Gru01}. Neutrons from the J\"ulich FRJ-2 reactor irradiated a $^{235}$U target and the fission fragments were identified with a gas-filled on-line mass separator. Gamma-rays of isomeric transitions were recorded with Ge(Li)-diodes. No details were given in the paper and the results were summarized in a table. The measured half-life of the 0.83(10)~$\mu$s isomeric state and the subsequent $\gamma$-ray energies agree with the presently accepted level scheme.

\subsection{$^{99}$Y}\vspace{0.0cm}

The first observation of $^{99}$Y was described in ``The P$_{n}$ Values of the $^{235}$U(n$_{th}$,f) Produced Precursors in the Mass Chains 90, 91, 93-95, 99, 134 and 137-139,'' in 1975 by Asghar et al. \cite{1975Asg01}. $^{235}$U targets were irradiated with neutrons from the Grenoble high flux reactor. The Lohengrin mass separator was used to identify the fission fragments by measuring the mass-to-charge ratio as well as the energy distribution and $\beta$-ray activities. ``The half-life of $^{99}$Y from this study is more precise than the value given by Eidens et al.'' The reported half-life of 1.45(22)~s agrees with the presently accepted value of 1.470(7)~s. The quoted paper by Eidens et al. \cite{1970Eid01} did not measure the half-life directly but only estimated it to 0.8(7)~s.

\subsection{$^{100}$Y}\vspace{0.0cm}

The identification of $^{100}$Y was first reported by Pfeiffer et al. in their 1977 paper ``Gamma Spectroscopy of Some Short-Lived Fission Products with the Isotope Separator Lohengrin'' \cite{1977Pfe01}.  $^{235}$U targets were irradiated with neutrons from the Grenoble high flux reactor. The Lohengrin mass separator was used to identify the fission fragments by measuring the mass-to-charge ratio as well as the energy distribution and $\gamma$-ray spectra. ``With the separator Lohengrin two groups of gamma-lines have been observed on mass 100. These lines have been separated according to their half-life (T$_{1/2}$ = 0.8 $\pm$ 0.3 s and T$_{1/2}$ = 7.0 $\pm$ 0.5 s) and attributed respectively to the decays of $^{100}$Y and $^{100}$Zr.'' The reported half-life of 0.8(3)~s is in good agreement with the accepted half-life of 735(7)~ms.

\subsection{$^{101}$Y}\vspace{0.0cm}

The existence of $^{101}$Y was first reported in ``Rotational Structure and Nilsson Orbitals for Highly Deformed Odd-A Nuclei in the A~100 Region,'' in 1983 by Wohn et al. \cite{1983Woh01}. Fission fragments produced by neutron irradiation at the Brookhaven high-flux beam reactor were analyzed with the TRISTAN on-line mass separator facility. ``Using ... a high-temperature surface-ionization ion source, we have studied decays of $^{99}$Sr, $^{101}$Sr, $^{99}$Rb, and $^{101}$Y and found half-lives (in milliseconds) of 266$\pm$5, 121$\pm$6, 52$\pm$5, and 500$\pm4$50, respectively.'' This half-life is included in the weighted average of the accepted half-life of 0.45(2)~s. Earlier assignments of low-energy $\gamma$-transitions \cite{1971Hop01} were evidently incorrect.

\subsection{$^{102}$Y}\vspace{0.0cm}

$^{102}$Y was first identified by Shizuma et al. in the 1983 paper ``Levels in $^{102}$Zr Populated in the Decay of $^{102}$Y'' \cite{1983Shi01}. Neutrons from the J\"ulich reactor irradiated a $^{235}$U target and the fission fragments were identified with the gas-filled recoil separator JOSEF. Decay curves and $\gamma$-ray spectra were measured. ``A least-squares fit to the data assuming only one half-life gives T$_{1/2}$=0.36$\pm$0.04 s for [the] $^{102}$Y decay.'' This half-life is the currently accepted half-life.

\subsection{$^{103-105}$Y}\vspace{0.0cm}

In 1994, Bernas et al. published the discovery of $^{103}$Y, $^{104}$Y, and $^{105}$Y  in ``Projectile Fission at Relativistic Velocities: A Novel and Powerful Source of Neutron-Rich Isotopes Well Suited for In-Flight Isotopic Separation'' \cite{1994Ber01}. The isotopes were produced using projectile fission of $^{238}$U at 750 MeV/nucleon on a lead target at GSI, Germany. ``Forward emitted fragments from $^{80}$Zn up to $^{155}$Ce were analyzed with the Fragment Separator (FRS) and unambiguously identified by their energy-loss and time-of-flight.'' This experiment yielded 972, 94, and 12 counts of $^{103}$Y, $^{104}$Y, and $^{105}$Y, respectively.

\subsection{$^{106,107}$Y}\vspace{0.0cm}

$^{106}$Y and $^{107}$Y were discovered by Bernas et al. in 1997, as reported in ``Discovery and Cross-Section Measurement of 58 New Fission Products in Projectile-Fission of 750$\cdot$AMeV $^{238}$U'' \cite{1997Ber01}. The experiment was performed using projectile fission of $^{238}$U at 750~MeV/nucleon on a beryllium target at GSI in Germany. ``Fission fragments were separated using the fragment separator FRS tuned in an achromatic mode and identified by event-by-event measurements of $\Delta$E-B$\rho$-ToF and trajectory.'' During the experiment, individual counts for $^{106}$Y (112) and $^{107}$Y (21) were recorded.

\subsection{$^{108,109}$Y}\vspace{0.0cm}

The discovery of $^{108}$Y and $^{109}$Y was reported in the 2010 article ``Identification of 45 New Neutron-Rich Isotopes Produced by In-Flight Fission of a $^{238}$U Beam at 345 MeV/nucleon,'' by Ohnishi et al. \cite{2010Ohn01}. The experiment was performed at the RI Beam Factory at RIKEN, where the new isotopes were created by in-flight fission of a 345 MeV/nucleon $^{238}$U beam on a beryllium target. $^{108}$Y and $^{109}$Y were separated and identified with the BigRIPS superconducting in-flight separator. The results for the new isotopes discovered in this study were summarized in a table. 132 individual counts for $^{108}$Y and six counts for $^{109}$Y were recorded.

\section{Discovery of $^{78-112}$Zr}

Thirty-five zirconium isotopes from A = 78--112 have been discovered so far; these include 5 stable, 12 neutron-deficient and 18 neutron-rich isotopes. According to the HFB-14 model \cite{2007Gor01}, $^{121}$Zr should be the last odd-even particle stable neutron-rich nucleus while the even-even particle stable neutron-rich nuclei should continue through $^{134}$Zr. On the proton dripline three more isotopes could be particle-stable ($^{75-77}$Zr). In addition, $^{73}$Zr and $^{74}$Zr could still have half-lives longer than 10$^{-9}$~ns \cite{2004Tho01}. Thus, about 21 isotopes have yet to be discovered corresponding to 38\% of all possible zirconium isotopes.

\begin{figure}
	\centering
	\includegraphics[scale=.5]{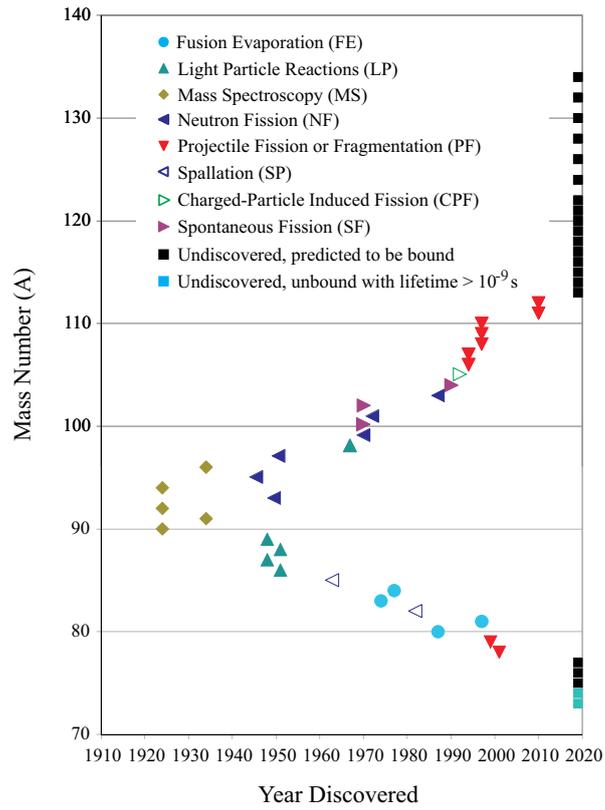}
	\caption{Zirconium isotopes as a function of time when they were discovered. The different production methods are indicated. The solid black squares on the right hand side of the plot are isotopes predicted to be bound by the HFB-14 model. On the proton-rich side the light blue squares correspond to unbound isotopes predicted to have lifetimes larger than $\sim 10^{-9}$~s.}
\label{f:year-zr}
\end{figure}

Figure \ref{f:year-zr} summarizes the year of first discovery for all zirconium isotopes identified by the method of discovery. The range of isotopes predicted to exist is indicated on the right side of the figure. The radioactive zirconium isotopes were produced using fusion evaporation reactions (FE), light-particle reactions (LP), neutron induced fission (NF), charged-particle induced fission (CPF), spontaneous fission (SF), spallation reactions (SP), and projectile fragmentation or fission (PF). The stable isotopes were identified using mass spectroscopy (MS). Light particles also include neutrons produced by accelerators. In the following, the discovery of each zirconium isotope is discussed in detail.

\subsection{$^{78}$Zr}\vspace{0.0cm}

In the 2001 article ``Synthesis and Halflives of Heavy Nuclei Relevant for the rp-Process,'' Kienle et al. first reported the existence of $^{78}$Zr \cite{2001Kie01}. A 1~A$\cdot$GeV $^{112}$Sn beam from the synchrotron SIS at GSI, Germany, bombarded a beryllium target. $^{78}$Zr was identified by the FRS fragment separator by both magnetic deflection and by measuring the energy loss and time-of-flight of the fragments. ``The spectra show the previously unobserved T = $-$1 nuclei $^{76}$Y (2 events) and $^{78}$Zr (one event) and demonstrate the absence of $^{81}$Nb...'' The observation of only one $^{78}$Zr event should be considered tentative until it is confirmed by an independent measurement.

\subsection{$^{79}$Zr}\vspace{0.0cm}

$^{79}$Zr was first reported in ``Observation of the Z = N + 1 Nuclei $_{39}^{77}$Y, $_{40}^{79}$Zr, and $_{42}^{83}$Mo'' in 1999 by Janas et al. \cite{1999Jan01}. At GANIL, France, nickel targets were bombarded with a 60 MeV/nucleon $^{92}$Mo beam. $^{79}$Zr was separated with the LISE3 spectrometer and the kinetic energy, energy loss, and time-of-flight were measured. ``The projections of the T$_{z}$ = 0 and $-1/2$ species onto the Z axis are presented in [the figure] and clearly show the presence of the even-Z, Z = N + 1 nuclei, $^{75}_{38}$Sr, $^{79}_{40}$Zr, and $^{83}_{42}$Mo in our spectra.''

\subsection{$^{80}$Zr}\vspace{0.0cm}

$^{80}$Zr was discovered by Lister et al. in their 1987 paper ``Gamma Radiation from the N = Z Nucleus $^{80}_{40}$Zr$_{40}$'' \cite{1987Lis01}. A magnesium target was bombarded with a 190 MeV $^{58}$Ni beam and $^{80}$Zr was produced in the fusion-evaporation reaction $^{24}$Mg($^{58}$Ni,2n). The residues were separated with the Daresbury recoil separator and $\gamma$ rays were detected with an array of bismuth germanate-shielded germanium detectors. ``Coincidence measurements between the isolated isotope and prompt gamma radiation allowed the identification of decays from low-lying states at E = 290 and 828 keV which indicate that $^{80}$Zr has an extremely large quadrupole deformation of $\beta_2 \sim$ 0.4.''

\subsection{$^{81}$Zr}\vspace{0.0cm}

In 1997, the first observation of $^{81}$Zr was published in ``Half-Lives of the T$_{z}$=1/2 Series Nuclei, $^{81}$Zr and $^{85}$Mo'' by Huang et al. \cite{1997Hua01}. At the Heavy-Ion Research Facility in Lanzhou, a 170 MeV $^{32}$S beam bombarded an enriched $^{58}$Ni target. $\gamma$-rays were measured with two coaxial HPGe detectors, while protons were detected with a Si(Au) surface barrier detector. ``By using the least squares method to fit the time spectrum, a half-life of 5.3$\pm$0.5 s is obtained.'' This half-life agrees with the accepted half-life of 5.5(4)~s. Earlier reports of half-lives of 7-15~min \cite{1965Zai01} and 15$\pm$5~s \cite{1982Del01} were evidently incorrect.

\subsection{$^{82}$Zr}\vspace{0.0cm}

The 1982 paper ``Selective On-Line Separation of New Ta, Zr and Sr Isotopes'' by Liang et al. described the discovery $^{82}$Zr \cite{1982Lia01}. A 280 MeV $^{3}$He beam from the IPN-Orsay synchrocyclotron was used to bombard SrF$_3$ powder and rolled molybdenum foils. $^{82}$Zr was identified with the ISOCELE-2 on-line mass separator. ``The half-life of 32$\pm$5~s differs from the previous value 9.5 mn and a tentatively proposed value 2.5 mn.'' The half-life presented in this paper has been adopted by the ENSDF evaluators as the currently accepted value for $^{82}$Zr based on the more direct method relative to earlier reports of half-lives of 9.5(35)~min \cite{1965Zai01} and 2.5~min \cite{1982Del01}.

\subsection{$^{83}$Zr}\vspace{0.0cm}

The first recorded observation of $^{83}$Zr was presented in ``Das Neue Isotop $^{83}$Zr'' by Kaba and Miyano in 1974 \cite{1974Kab01}. Enriched $^{54}$Fe targets were bombarded with a 100 MeV $^{32}$S beam from the Munich MP tandem accelerator. $^{83}$Zr was identified by measuring $\gamma$-ray spectra with Ge(Li) detectors following chemical separation. ``Die Halbwertszeiten der sieben gefundenen Gammalinien entsprechen genau der Anstiegzeit des $^{83m}$Y. Auf Grund dessen k\"onnen Gammalinien mit T$_{1/2}$ = 42 sec dem Mutterkern von $^{83m}$Y, n\"amlich $^{83}$Zr zugeordnet werden.'' [ The half-lives of the seven observed $\gamma$-rays correspond exactly to the time for the ingrowing $^{83m}$Y. For this reason the $\gamma$-lines with T$_{1/2}$ = 42~s were attributed to $^{83}$Zr, the mother nucleus of $^{83m}$Y.] This half-life agrees with the presently accepted value of 41.6(24)~s. A similar half-life of 0.7~min was attributed to either $^{83}$Zr or $^{80}$Y \cite{1971Dor01}. Also, a previous observation of a half-life of 5$-$10~min assigned to $^{83}$Zr was evidently incorrect \cite{1965Zai01}.

\subsection{$^{84}$Zr}\vspace{0.0cm}

``Investigation of Neutron Deficient Zr and Nb Nuclei with Heavy Ion Induced Compound Reactions,'' published in 1977 by Korschinek et al., described the first observation of $^{84}$Zr \cite{1977Kor01}. $^{32}$S and $^{12}$C beams were used to bombard targets of $^{54}$Fe, $^{58}$Ni, and $^{74}$Se at the Munich MP tandem accelerator. The $\gamma$-ray emissions of the reaction products were measured with a set of coaxial Ge(Li) detectors. ``The statistics in the case of $^{84}$Zr are poorer. A $\gamma$ cascade with the transitions at 540, 722.8 and 874.4 keV could be identified.''  The currently adopted value for the half-life is 25.8(5)~min. Previously half-life measurements of 16(4)~min \cite{1965Zai01} and 5.0(5)~min \cite{1971Yu01} turned out to be incorrect.

\subsection{$^{85}$Zr}\vspace{0.0cm}

In the 1963 article ``Neutron-deficient Isotopes of Yttrium and Zirconium,'' Butement and Briscoe reported the discovery of $^{85}$Zr \cite{1963But01}. A beam of 230 MeV protons irradiated targets of dry yttrium oxide and strontium oxide. Decay curves were measured with end-window Geiger counters and $\gamma$-ray spectra were recorded with a NaI(T1) crystal following chemical separation. ``A new activity with a half-life of 6 min has been assigned to $^{85}$Zr,...'' This half-life is close to the currently accepted half-life of 7.86(4)~min.

\subsection{$^{86}$Zr}\vspace{0.0cm}

Hyde and O'Kelley first identified $^{86}$Zr in their paper, ``Radiochemical and Spectrometer Studies of Several Neutron-Deficient Zirconium Isotopes and Their Decay Products,'' in 1951 \cite{1951Hyd01}. Thin strips of niobium metal were bombarded with 100 MeV protons from the Berkeley 184-inch synchrocyclotron. The spallation products were separated through rapid radiochemistry and studied with a $\beta$-ray spectrometer. ``Zr$^{86}$ is a 17$\pm$2-hour orbital electron capturing isotope decaying into Y$^{86}$, which in turn disintegrates into stable Sr$^{86}$ with a half-life of 14.6$\pm$0.2 hours by the emission of positrons.'' This half-life agrees with the presently accepted value of 16.5(1)~h.

\subsection{$^{87}$Zr}\vspace{0.0cm}

$^{87}$Zr was first observed by Robertson et al. in ``Radioactive Y$^{84}$, Y$^{88}$, and Zr$^{87}$,'' published in 1949 \cite{1949Rob01}. A beam of 20 MeV $\alpha$-particles bombarded enriched targets of Sr$^{84}$O and Sr$^{86}$O. The positron activity was measured with a calibrated spectrometer counter. ``From the percent composition of the strontium in each of the two samples, one is lead to conclude that the 2.0-hour activity is caused by the strontium 84 and should be assigned to zirconium 87. The reaction is Sr$^{84}$($\alpha$,n)Zr$^{87}$.'' This half-life is close to the accepted half-life of 1.68(1)~h.

\subsection{$^{88}$Zr}\vspace{0.0cm}

Hyde and O'Kelley first identified $^{88}$Zr in their paper, ``Radiochemical and Spectrometer Studies of Several Neutron-Deficient Zirconium Isotopes and Their Decay Products,'' in 1951 \cite{1951Hyd01}. Thin strips of niobium metal were bombarded with 100 MeV protons from the Berkeley 184-inch synchrocyclotron. The spallation products were separated through rapid radiochemistry and studied with a $\beta$-ray spectrometer. ``Preliminary evidence for Zr$^{88}$ is presented which indicates that this isotope decays, by the capture of orbital electrons with a half-life of the order of 150 days, into 105-day Y$^{88}$.'' This approximate value for the half-life is within a factor of two of the presently accepted value of 83.4(3)~d.

\subsection{$^{89}$Zr}\vspace{0.0cm}

Sagane et al. reported the first observation of $^{89}$Zr in the 1938 paper ``A Preliminary Report on the Radioactivity Produced in Y, Zr, and Mo'' \cite{1938Sag02}. Fast neutrons produced by 3 MeV deuterons on lithium and beryllium at the cyclotron of the Institute of Physical and Chemical Research in Tokyo, Japan, irradiated zirconium targets. Positrons were detected following chemical separation. No further details were given and a half-life of 70~h was listed for $^{89}$Zr in a table. This value is close to the accepted half-life of 78.41(12)~h.

\subsection{$^{90}$Zr}\vspace{0.0cm}

In 1924 Aston identified the stable isotope $^{90}$Zr in ``The Mass Spectra of Zirconium and Some Other Elements'' \cite{1924Ast02}. The Cavendish laboratory mass spectrograph was used to separate accelerated anode rays and identify their mass. ``Zirconium gives mass lines 90, 92, 94, and a doubtful one at (96), with relative intensities, very roughly, 10, 2, 4, (1) respectively.'' These values are in fair agreement with the presently adopted relative abundances of 51.45\%, 17.15\%, 17.38\% and 2.8\%, respectively.

\subsection{$^{91}$Zr}\vspace{0.0cm}

$^{91}$Zr was discovered by Aston in ``Constitution of Hafnium and other Elements'' in 1934 \cite{1934Ast02}. The stable isotope was identified with an anode discharge tube installed at the Cavendish Laboratory mass spectrograph. ``New mass-spectra obtained from zirconium not only show an additional and fairly abundant isotope 91, hitherto overlooked owing to insufficient resolution, but also confirm the presence of the very rare and previously doubtful constituent 96, which is of particular interest as it forms with molybdenum and ruthenium the lightest known isobaric triplet.''

\subsection{$^{92}$Zr}\vspace{0.0cm}

In 1924 Aston identified the stable isotope $^{92}$Zr in ``The Mass Spectra of Zirconium and Some Other Elements'' \cite{1924Ast02}. The Cavendish laboratory mass spectrograph was used to separate accelerated anode rays and identify their mass. ``Zirconium gives mass lines 90, 92, 94, and a doubtful one at (96), with relative intensities, very roughly, 10, 2, 4, (1) respectively.'' These values are in fair agreement with the presently adopted relative abundances of 51.45\%, 17.15\%, 17.38\% and 2.8\%, respectively.

\subsection{$^{93}$Zr}\vspace{0.0cm}

Steinberg and Glendenin published evidence for $^{93}$Zr in their 1950 paper ``A Long-Lived Zirconium Activity in Fission'' \cite{1950Ste01}. Uranium was irradiated in a pile for ten months and $^{93}$Zr was chemically separated four years after the end of the irradiation. The activity was then measured with a thin end-window proportional counter and an internal Geiger counter. ``In the present investigation, we have isolated from uranium fission a zirconium activity of $\sim5\times10^6$-yr. half-life, emitting beta-rays of 60$\pm$5 kev maximum energy, which is probably Zr$^{93}$... Thus, the true half-life of Zr$^{93}$ should lie in the range 1.5 to 8.5$\times$10$^6$ yr.'' This estimate agrees with the presently adopted value of 1.53(10)$\times$10$^6$ years. A previously reported half-life of 63(5)~d \cite{1940Sag01} was evidently incorrect.

\subsection{$^{94}$Zr}\vspace{0.0cm}

In 1924 Aston identified the stable isotope $^{94}$Zr in ``The Mass Spectra of Zirconium and Some Other Elements'' \cite{1924Ast02}. The Cavendish laboratory mass spectrograph was used to separate accelerated anode rays and identify their mass. ``Zirconium gives mass lines 90, 92, 94, and a doubtful one at (96), with relative intensities, very roughly, 10, 2, 4, (1) respectively.'' These values are in fair agreement with the presently adopted relative abundances of 51.45\%, 17.15\%, 17.38\% and 2.8\%, respectively.

\subsection{$^{95}$Zr}\vspace{0.0cm}

$^{95}$Zr was first observed by Grummitt and Wilkinson in their 1946 paper ``Fission Products of U$^{235}$'' \cite{1946Gru01}. The isotope was obtained in thermal neutron fission of $^{235}$U. Activities were measured with a $\beta^-$-spectrometer following chemical separation. ``Several previously unreported isotopes have been observed during the course of the present work... Fission yields of these and the following $\beta^{-}$ active isotopes were measured: Zr$^{95}$ (65 days, 0.5 MeV).'' This half-life agrees with the presently accepted value of 64.032(6)~d. A previously reported half-life of 17(1)~h \cite{1940Sag01} was evidently incorrect.

\subsection{$^{96}$Zr}\vspace{0.0cm}

$^{96}$Zr was discovered by Aston in ``Constitution of Hafnium and other Elements'' in 1934 \cite{1934Ast02}. The stable isotope was identified with an anode discharge tube installed at the Cavendish Laboratory mass spectrograph. ``New mass-spectra obtained from zirconium not only show an additional and fairly abundant isotope 91, hitherto overlooked owing to insufficient resolution, but also confirm the presence of the very rare and previously doubtful constituent 96, which is of particular interest as it forms with molybdenum and ruthenium the lightest known isobaric triplet.''

\subsection{$^{97}$Zr}\vspace{0.0cm}

$^{97}$Zr was identified by Katcoff and Finkle in the 1945 paper ``Energies of Radiations of 17h Zr$^{97}$ and 75m Nb$^{97}$'' \cite{1951Kat02}. $^{239}$Pu was irradiated at the Argonne Heavy-Water Pile and $\beta$- and $\gamma$-rays were measured following chemical separation. ``By subtracting the aluminum absorption curve of 75m Nb$^{97}$ from the curve for the mixture, the absorption curve for pure Zr$^{97}$ was obtained.'' Katcoff and Finkle assumed the half-life of 17~h to be known, however, we could not find any previous publication assigning this half-life to $^{97}$Zr. It had been a known zirconium activity with no mass assignment \cite{1940Gro01,1941Hah02,1951Bra01}. Sagane et al. had incorrectly assigned it to $^{95}$Zr \cite{1940Sag01}. They also incorrectly assigned a half-life of 6(1)~m to $^{97}$Zr. Technically, the first refereed publication was by Burgus et al. in 1950 \cite{1950Bur01}, however, as participants of the Plutonium Project, they had access and gave credit to the work of Katcoff and Finkle.

\subsection{$^{98}$Zr}\vspace{0.0cm}

H\"ubenthal was the first to identify $^{98}$Zr in 1967 as reported in ``La cha\^ine isobarique de masse 98: $^{98}$Zr $\rightarrow ^{98}$Nb $\rightarrow ^{98}$Mo dans la fission de l'uranium 235 par neutrons thermiques'' \cite{1967Hub02}. Thermal neutrons from the Grenoble Silo\'e reactor irradiated a $^{235}$U target. Beta-decay spectra were recorded with a plastic scintillator following chemical separation. ``En [figure] (b) la courbe exp\'erimentale a \'et\'e d\'ecompos\'ee, soit en utilisant les composantes trouv\'ees en [figure] (a), soit en prolongeant la partie comprise entre t=10~mn et t=20~mn, les deux m\'ethodes donnent le m\^eme r\'esultat, on voit sur plus d'une d\'ecade une p\'eriode de 31$\pm$3~s que nous attribuons \`a $^{98}$Zr.'' [In [figure](b) the experimental curve has been decomposed, either by using components found in [figure] (a) or by extending the part between t=10~min and t=20~min - the two methods give the same result - a half-life of 31$\pm$3~s can be seen over more than one decay period which we attribute to Zr.'' This half-life agrees with the presently adopted value of 30.7(4)~s. Previously, a 1~min half-life was observed in zirconium but no firm assignment to $^{98}$Zr was made \cite{1960Ort01}.

\subsection{$^{99}$Zr}\vspace{0.0cm}

In 1970, Eidens et al. described the first observation of $^{97}$Y in ``On-Line Separation and Identification of Several Short-Lived Fission Products: Decay of $^{84}$Se, $^{91}$Kr, $^{97}$Y, $^{99}$Nb, $^{99}$Zr, $^{100,101}$Nb and $^{101}$Zr'' \cite{1970Eid01}. Neutrons from the J\"ulich FRJ-2 reactor irradiated a $^{235}$U target and the fission fragments were identified with a gas-filled on-line mass separator. Beta-$\gamma$- and $\gamma$-$\gamma$-coincidences were recorded. ``A 468$\pm$3 keV line and a 548$\pm$3 keV line were detected as members of a $\gamma$-$\gamma$ cascade in the coincidence investigations. They were assigned to $^{99}$Zr'' The extracted half-life of 2.4(1)~s agrees with the currently adopted value of 2.1(1)~s. Previously only an upper limit of 1.6~s \cite{1963Tro01} was reported for the half-life of $^{99}$Zr and a measurement of 35(5)~s \cite{1960Ort01} was evidently incorrect.

\subsection{$^{100}$Zr}\vspace{0.0cm}

Evidence for $^{100}$Zr was observed by Cheifetz et al. in the 1970 paper ``Experimental Information Concerning Deformation of Neutron Rich Nuclei In the A $\sim$ 100 Region'' \cite{1970Che01}. Fission fragments from the spontaneous fission of $^{252}$Cf were observed in coincidence with X- and $\gamma$-rays measured in Ge(Li) detectors. Several isotopes were identified and the observed $\gamma$-rays were listed in a table. For $^{100}$Zr, levels at 212.7 (2$^+$), 564.8 (4$^+$), and 1062.7 keV (6$^+$) were reported. These levels agree with the presently accepted level scheme. An earlier estimate of 1.0(9)~s \cite{1970Eid01} for the half-life of $^{100}$Zr was incorrect.

\subsection{$^{101}$Zr}\vspace{0.0cm}

$^{101}$Zr was first correctly identified by Trautmann et al. in the 1972 paper ``Identification of Short-Lived Isotopes of Zirconium, Niobium, Molybdenum, and Technetium in Fission by Rapid Solvent Extraction Techniques'' \cite{1972Tra01}. $^{235}$U and $^{239}$Pu targets were irradiated with thermal neutrons at the Mainz Triga reactor. Following chemical separation, $\gamma$-ray spectra were recorded with a Ge(Li) detector. ``2.0-sec $^{101}$Zr. An average half-life of 2.0$\pm$0.3 sec was obtained for this nuclide from the growth of its $^{101}$Nb daughter, and from its genetic relationship with the 14-min. $^{101}$Tc.'' This value is currently included as part of the weighted average of the accepted half-life for $^{101}$Zr, 2.3(1)~s. Trautmann et al. indicated that an earlier report of a 3.3(6)~s half-life for $^{101}$Zr \cite{1970Eid01} was misidentified and should be assigned to $^{102}$Nb.

\subsection{$^{102}$Zr}\vspace{0.0cm}

In ``A Study of the Low-Energy Transitions Arising from the Prompt De-Excitation of Fission Fragments,'' published in 1970, Watson et al. reported the first observation of $^{102}$Zr \cite{1970Wat01}. Fission fragments from the spontaneous fission of $^{252}$Cf were measured in coincidence with X-rays and conversion electrons. In a table, the energy (153~keV) and half-life (1.7~ns) of the first excited state were identified correctly as they agree with presently accepted values.

\subsection{$^{103}$Zr}\vspace{0.0cm}

``Nuclear Structure Effects in the Mass Region Around A = 100, Derived from Experimental Q$_{\beta}$-Values,'' published in 1987 by Graefenstedt et al., reported the first identification of $^{103}$Zr \cite{1987Gra01}. A $^{239}$Pu target was used at the Lohengrin Mass Separator Facility in Grenoble. Beta-decay energies measured with a plastic scintillator telescope were recorded in coincidence with  $\gamma$-rays measured in a large Ge-detector. ``Five endpoints of $\beta$-transitions to excited levels in $^{103}$Nb could be measured in the present investigation. From them, a consistent Q$_\beta$-value of 6945$\pm$85 keV is obtained.'' The reported half-life of 1.3~s is the presently adopted value of 1.3(1)~s as quoted from an unpublished report \cite{1980Sch02}.

\subsection{$^{104}$Zr}\vspace{0.0cm}

The first clean identification of $^{104}$Zr was described in ``New Neutron-Rich Nuclei $^{103,104}$Zr and the A$\sim$100 Region of Deformation,'' by Hotchkis et al. in 1990 \cite{1990Hot01}. The $^{104}$Zr level scheme was measured at the Argonne-Notre Dame $\gamma$-ray facility following the spontaneous fission of $^{248}$Cm with an array of ten bismuth-germanate-suppressed Ge detectors and fifty bismuth-germanate scintillators. ``We have for the first time determined partial decay schemes in the nuclei $^{103}$Zr and $^{104}$Zr and determined extensive new data on $^{100-102}$Zr.'' Levels at 140.3, 452.8, 926.5, and 1551.3 keV were presented in a level scheme. These levels agree with the presently adopted level scheme. Hotchkis et al. also claim the first observation of $^{103}$Zr. They were apparently not aware of the work by Graefenstedt et al. \cite{1987Gra01}. A previous measurement of a level at 251.7~keV was evidently incorrect \cite{1972Hop01}.

\subsection{$^{105}$Zr}\vspace{0.0cm}

The first observation of $^{105}$Zr was reported by \"Ayst\"o et al. in ``Discovery of Rare Neutron-Rich Zr, Nb, Mo, Tc and Ru Isotopes in Fission: Test of $\beta$ Half-life Predictions Very Far from Stability'' in 1992 \cite{1992Ays01}. At the Ion Guide Isotope Separator On-Line (IGISOL) in Jyv\"askyl\"a, Finland, targets of uranium were bombarded with 20 MeV protons. $\beta$ decays were measured with a planar Ge detector, while $\gamma$-rays were measured with a 50\% Ge detector located behind a thin plastic detector. ``Only a slight indication of $^{105}$Zr is seen in the A=105 spectrum. However, at this mass number we detected a 127.9-keV $\gamma$ transition in coincidence with $\beta$ particles giving further evidence for $^{105}$Zr.'' The measured half-life of 1000$^{+1200}_{-400}$~s is consistent with the presently accepted value of 0.6(1)~s.

\subsection{$^{106,107}$Zr}\vspace{0.0cm}

In 1994, Bernas et al. published the discovery of $^{106}$Zr and $^{107}$Zr in ``Projectile Fission at Relativistic Velocities: A Novel and Powerful Source of Neutron-Rich Isotopes Well Suited for In-Flight Isotopic Separation'' \cite{1994Ber01}. The isotopes were produced using projectile fission of $^{238}$U at 750 MeV/nucleon on a lead target at GSI, Germany. ``Forward emitted fragments from $^{80}$Zn up to $^{155}$Ce were analyzed with the Fragment Separator (FRS) and unambiguously identified by their energy-loss and time-of-flight.'' This experiment yielded 336 and 38 counts of $^{106}$Zr and $^{107}$Zr, respectively.

\subsection{$^{108-110}$Zr}\vspace{0.0cm}

$^{108}$Zr, $^{109}$Zr, and $^{110}$Zr were discovered by Bernas et al. in 1997, as reported in ``Discovery and Cross-Section Measurement of 58 New Fission Products in Projectile-Fission of 750$\cdot$AMeV $^{238}$U'' \cite{1997Ber01}. The experiment was performed using projectile fission of $^{238}$U at 750~MeV/nucleon on a beryllium target at GSI in Germany. ``Fission fragments were separated using the fragment separator FRS tuned in an achromatic mode and identified by event-by-event measurements of $\Delta$E-B$\rho$-ToF and trajectory.''  During the experiment, individual counts for $^{108}$Zr (336), $^{109}$Zr (47), and $^{110}$Zr (5) were recorded.

\subsection{$^{111,112}$Zr}\vspace{0.0cm}

The discovery of $^{111}$Zr and $^{112}$Zr was reported in the 2010 article ``Identification of 45 New Neutron-Rich Isotopes Produced by In-Flight Fission of a $^{238}$U Beam at 345 MeV/nucleon,'' by Ohnishi et al. \cite{2010Ohn01}. The experiment was performed at the RI Beam Factory at RIKEN, where the new isotopes were created by in-flight fission of a 345 MeV/nucleon $^{238}$U beam on a beryllium target. $^{111}$Zr and $^{112}$Zr were separated and identified with the BigRIPS superconducting in-flight separator. The results for the new isotopes discovered in this study were summarized in a table. Twenty-six individual counts for $^{111}$Zr and one count for $^{112}$Zr were recorded. The observation of only one $^{112}$Zr event should be considered tentative until it is confirmed by an independent measurement.

\section{Discovery of $^{82-115}$Nb}

For a long time the naming of the element niobium was controversial. It was discovered in 1801 by Hatchett who named it columbium. A committee of the Council of the International Association of Chemical Societies endorsed the name niobium in 1913 \cite{1914Cla01}. Columbian was continued to be used in the U.S. while in Europe niobium was used. In 1949, at the 15$^{th}$ Conference of the International Union of Chemistry, niobium was voted to be the accepted name \cite{1949UIC01}.

Thirty-four niobium isotopes from A = 82--115 have been discovered so far; these include 1 stable, 11 neutron-deficient and 22 neutron-rich isotopes.
According to the HFB-14 model \cite{2007Gor01}, $^{121}$Nb should be the last odd-odd particle stable neutron-rich nucleus while the odd-even particle stable neutron-rich nuclei should continue through $^{134}$Y. The proton dripline has most likely been reached at $^{82}$Nb, because $^{81}$Nb was shown to be particle unstable \cite{1999Jan01}. However, five more isotopes $^{76-81}$Nb could possibly still have half-lives longer than 10$^{-9}$~ns \cite{2004Tho01}. Thus, about 22 isotopes have yet to be discovered corresponding to 39\% of all possible niobium isotopes.

\begin{figure}
	\centering
	\includegraphics[scale=.5]{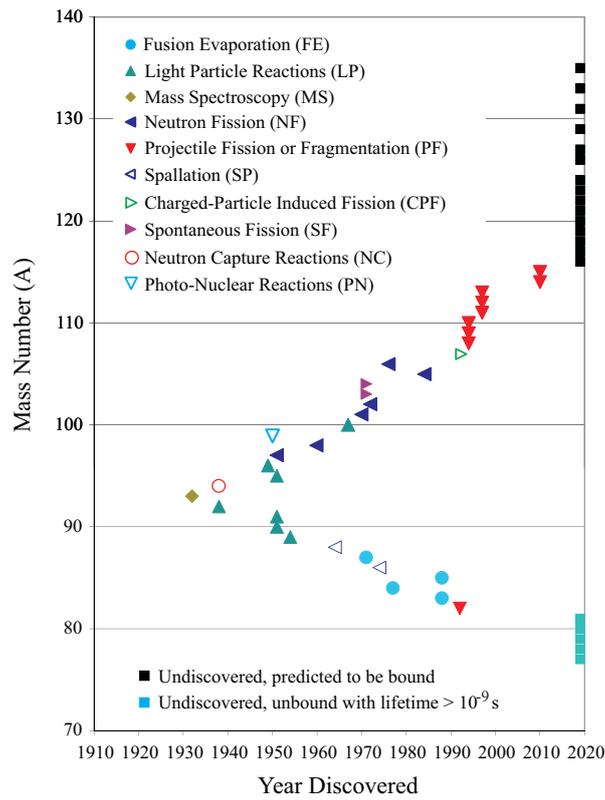}
	\caption{Niobium isotopes as a function of time when they were discovered. The different production methods are indicated. The solid black squares on the right hand side of the plot are isotopes predicted to be bound by the HFB-14 model. On the proton-rich side the light blue squares correspond to unbound isotopes predicted to have lifetimes larger than $\sim 10^{-9}$~s.}
\label{f:year-nb}
\end{figure}

Figure \ref{f:year-nb} summarizes the year of first discovery for all niobium isotopes identified by the method of discovery. The range of isotopes predicted to exist is indicated on the right side of the figure. The radioactive niobium isotopes were produced using fusion evaporation reactions (FE), light-particle reactions (LP), neutron induced fission (NF), charged-particle induced fission (CPF), spontaneous fission (SF), neutron capture (NC), photo-nuclear (PN), spallation reactions (SP), and projectile fragmentation or fission (PF). The stable isotope was identified using mass spectroscopy (MS). Light particles also include neutrons produced by accelerators. In the following, the discovery of each niobium isotope is discussed in detail.

\subsection{$^{82}$Nb}\vspace{0.0cm}

The discovery of $^{82}$Nb is credited to Yennello et al. with their 1992 paper ``New Nuclei Along the Proton-Drip Line Near Z = 40,'' \cite{1992Yen01}.  At the National Superconducting Cyclotron Laboratory at Michigan State University, a 70~MeV/A $^{92}$Mo beam was produced by the K1200 cyclotron and impinged on a $^{58}$Ni target. $^{82}$Nb was identified with the A1200 fragment analyzer by measuring the time-of-flight and energy loss of the fragments. ``The mass spectra for residues with Z from 39 to 44 are shown in [the figure] with the new isotopes marked by arrows. Although both $^{84}$Mo and $^{86}$Mo have been previously observed, no reference to the identification of $^{85}$Mo was found. The other new isotopes observed in this study are $^{78}$Y, $^{82}$Nb, $^{86}$Tc, and $^{89,90}$Ru.''

\subsection{$^{83}$Nb}\vspace{0.0cm}

$^{83}$Nb was discovered by Kuroyanagi et al., as reported in their 1988 paper, ``New Neutron-Deficient Isotopes $^{83}$Nb and $^{85}$Nb'' \cite{1988Kur01}. A 95 MeV $^{28}$Si beam from the Kyushu University tandem accelerator bombarded an enriched $^{58}$Ni target and $^{83}$Nb was formed in the fusion evaporation reaction $^{58}$Ni($^{28}$Si,p2n). Gamma- and $\beta$-rays were detected with a Ge detector and plastic scintillator following the irradiation. ``The activity with the half-life of 4.1 sec is undoubtedly assigned to a previously unidentified nuclide of $^{83}$Nb, because of the appearance of two gamma-rays of which the energies agree with the transition energies from first and second excited states of $^{83}$Zr, and cascade relations of the gamma- and beta-rays.'' The half-life of 4.1(3)~s is currently used as the sole value for the accepted half-life.

\subsection{$^{84}$Nb}\vspace{0.0cm}

``Investigation of Neutron Deficient Zr and Nb Nuclei with Heavy Ion Induced Compound Reactions,'' published in 1977 by Korschinek et al., described the first observation of $^{84}$Nb \cite{1977Kor01}. $^{32}$S beams with energies between 93 and 120 MeV were used to bombard enriched $^{58}$Ni targets at the Munich MP tandem accelerator. $^{84}$Nb was formed in the fusion evaporation reaction $^{58}$Ni($^{32}$S,np$\alpha$) and identified by the $\gamma$-ray spectra measured with a set of coaxial Ge(Li) detectors. ``The ground state band up to $J^\pi$ = (4$^+$) in $^{84}$Zr and up to $J^\pi$ = (8$^+$) in $^{86}$Zr were observed in the present $\gamma$ spectra of the residual activities of the irradiation $^{58}$Ni + $^{32}$S with half-lives of T$_{1/2}$ = 12$\pm$3~s and T$_{1/2}$ = 80$\pm$12~s, respectively. These activities are believed to originate from the decay of the new isotopes $^{84}$Nb and $^{86}$Nb, respectively.'' The half-life of 9.8(9)~s is included in the weighted average for the presently accepted value.

\subsection{$^{85}$Nb}\vspace{0.0cm}

$^{85}$Nb was discovered by Kuroyanagi et al., as reported in their 1988 paper, ``New Neutron-Deficient Isotopes $^{83}$Nb and $^{85}$Nb'' \cite{1988Kur01}. A 105 MeV $^{32}$S beam from the Kyushu University tandem accelerator bombarded an enriched $^{58}$Ni target and $^{85}$Nb was formed in the fusion evaporation reaction $^{58}$Ni($^{32}$S,$\alpha$p). Gamma- and $\beta$-rays were detected with a Ge detector and plastic scintillator following the irradiation. ``The activity with the half-life of 20.9 sec is uniquely attributed to a previously unknown isotope of $^{85}$Nb, because the energy value of 50.1 keV gamma-ray appearing in the present experiments agrees with the energy of the first excited level of 50.07$\pm$0.03 keV which has been established in the in-beam study of $^{85}$Zr.'' The half-life of 20.9(7)~s is currently used as the sole value for the accepted half-life. An earlier report of a 2.3(3)~min half-life \cite{1982Del01} was evidently incorrect.

\subsection{$^{86}$Nb}\vspace{0.0cm}

The observation of $^{86}$Nb was reported in ``Decay of $^{87}$Nb. The New Isotope $^{86}$Nb,'' by Votsilka et al. in 1974 \cite{1974Vot01}. A beam of 600 MeV protons bombarded a silver target at the Joint Institute for Nuclear Studies. $^{86}$Nb was identified by measuring $\gamma$-ray spectra with a Ge(Li) detector following chemical separation. ``Analysis of the intensity of the $\gamma$ transition at 243 kev in$^{86}$Zr (16.5 hr) in specimens successively separated out of the niobium fraction showed that the half-life of the parent $^{86}$Nb was 1.6$\pm$0.7 min.'' This half-life agrees with the currently adopted value of 88(1)~s.

\subsection{$^{87}$Nb}\vspace{0.0cm}

In ``Decay Scheme Studies of Short-Lived Isotopes of 69 $\leqq$ A $\leqq$ 88 Produced by Heavy-Ion Bombardment,'' published in 1971, Doron and Blann described the first observation of $^{87}$Nb \cite{1971Dor01}. Natural nickel and enriched $^{58}$Ni targets were bombarded with 85 to 100 MeV $^{32}$S beams at the University of Rochester MP Tandem Van de Graaff accelerator. $^{87}$Nb was identified by measuring half-lives, excitation functions and $\gamma$-rays. ``According to the calculated excitation functions, the relative cross sections for the 3.5 min  activity and the observed growth of the $^{87m}$Y isomeric transition, this activity could possibly be due to either the decay of $^{87}$Nb or of $^{87}$Mo. The fact that the  predicted cross section for $^{87}$Mo production by the $^{59}$Co($^{32}$S,p3n)$^{87}$Mo reaction is smaller by a factor of 100 than the $^{59}$Co($^{32}$S,2p2n)$^{87}$Nb cross section suggests that this activity is due to the decay of $^{87}$Nb.'' The measured halflife of 3.5(2)~min is part of the weighted average of the accepted half-life.

\subsection{$^{88}$Nb}\vspace{0.0cm}

$^{88}$Nb was discovered by Korteling and Hyde and is reported in their 1964 paper ``Interaction of High-Energy Protons and Helium Ions with Niobium'' \cite{1964Kor01}. Niobium foils were bombarded with 240-720 MeV protons and 320-880 MeV helium ions from the Lawrence Radiation Laboratory 184-in. synchrocyclotron. Beta- and $\gamma$-rays were measured following chemical separation. ``During the course of these experiments, a 15-min positron activity was observed in the niobium fraction and it was established that this activity was the parent of Zr$^{88}$. This conclusion was confirmed by preparation of the Nb$^{88}$ activity by the reaction of carbon nuclei with a bromine target in the Berkeley heavy ion linear accelerator.'' The half-life agrees with the presently accepted value of 14.5(1)~min. Butement and Qaim reported a half-life of 21 min only six months earlier \cite{1964But03}, however, later their data were found to be incorrect \cite{1966Fle01}.

\subsection{$^{89}$Nb}\vspace{0.0cm}

Diamond reported the discovery of $^{89}$Nb in ``New Isotopes of Niobium: Nb$^{89}$ and Nb$^{89m}$'' in 1954 \cite{1954Dia02}. Niobium and zirconium metal foils were irradiated with 60 MeV protons from the Harvard synchrocyclotron. $^{89}$Nb was identified by measuring the activity with an end-window Geiger counter following chemical separation. ``A new 1.9$\pm$0.3-hour activity was found in niobium fractions isolated from proton irradiated niobium and zirconium metal foils. By identification of the 78-hour daughter activity, the mass assignment has been made to Nb$^{89}$.'' This half-life is consistent with the presently accepted value of 2.03(7)~h.

\subsection{$^{90,91}$Nb}\vspace{0.0cm}

Jacobson and Overstreet identified $^{90}$Nb and $^{91}$Nb in ``Mass Assignment of the Chain 65d Zr$^{95}$ - 35d Nb$^{95}$ and Notes on Other Niobium Activities,'' in 1951 as a part of the Plutonium Project Series \cite{1951Jac01}. Deuterons and fast neutrons produced with the Berkeley cyclotron bombarded molybdenum and zirconium targets and niobium targets, respectively. Decay and absorption curves were measured following chemical separation. ``In addition, evidence was obtained that the 60d Nb has a mass assignment of 91, clarifying certain previous transmutations studies, and that the 18h Nb positron emitter is Nb$^{90}$.'' These half-lives are close to the currently adopted values of 14.60(5)~h and 60.86(22)~d for $^{90}$Nb and $^{91}$Nb, respectively. The half-life for $^{91}$Nb corresponds to an isomeric state.

\subsection{$^{92}$Nb}\vspace{0.0cm}

Sagane et al. described the discovery of $^{92}$Nb in the 1938 paper ``A Preliminary Report on the Radioactivity Produced in Y, Zr, and Mo'' \cite{1938Sag03}. Fast neutrons produced by bombarding lithium with 3 MeV deuterons from the Tokyo cyclotron were used to irradiate niobium samples. Beta-ray activities were measured following chemical separation. The half-life of $^{92}$Nb was reported to be 11(1)~d in a table. This value agrees well with the half-life of 10.15(2)~d for an isomeric state in $^{92}$Nb.

\subsection{$^{93}$Nb}\vspace{0.0cm}

In 1932 Aston discovered the only stable isotope of niobium, $^{93}$Nb, as reported in ``Constitution of Tantalum and Niobium'' \cite{1932Ast01}. Niobium penta-fluoride was volatilised in the discharge tube of the Cavendish mass spectrometer. ``Niobium behaved in exactly the same way, giving a single line at 93 and fluorides at 102, 121...''

\subsection{$^{94}$Nb}\vspace{0.0cm}

Sagane et al. described the discovery of $^{94}$Nb in the 1938 paper ``A Preliminary Report on the Radioactivity Produced in Y, Zr, and Mo'' \cite{1938Sag03}. Neutrons produced by bombarding beryllium with 3 MeV deuterons from the Tokyo cyclotron and slowed down with paraffin were used to irradiate niobium samples. Beta-ray activities were measured to identify the isotope which was produced by neutron capture on stable $^{91}$Nb. The half-life of $^{92}$Nb was reported to be 7.5(5)~m in a table. ``The 7.5m period is in good agreement with Pool, Cork and Thornton.'' This value agrees well with the half-life of 6.263(4)~m for an isomeric state in $^{94}$Nb. The work by Pool et al. referred to by Sagane et al. reported a half-life of 7.3~m but was not able to make a unique mass assignment.

\subsection{$^{95}$Nb}\vspace{0.0cm}

Jacobson and Overstreet identified $^{95}$Nb in ``Mass Assignment of the Chain 65d Zr$^{95}$ - 35d Nb$^{95}$ and Notes on Other Niobium Activities,'' in 1951 as a part of the Plutonium Project Series \cite{1951Jac01}. Deuterons and fast neutrons produced with the Berkeley cyclotron bombarded molybdenum and zirconium targets and niobium targets, respectively. Decay and absorption curves were measured following chemical separation. ``The 35d Nb in question grows from a zirconium parent with a half-life of about 65 days; therefore it cannot be Nb$^{93}$ and must be Nb$^{95}$.'' The half-life agrees with the currently adopted value of 34.991(6)~d. A previously reported half-life of 62(5)~min \cite{1940Sag01} was evidently incorrect.

\subsection{$^{96}$Nb}\vspace{0.0cm}

$^{96}$Nb was identified correctly by Kundu and Pool in their 1949 paper ``Columbium 96'' \cite{1949Kun01}. Enriched $^{90}$Zr, $^{91}$Zr, $^{92}$Zr, and $^{96}$Zr targets were bombarded with 5 MeV protons and 10 MeV deuterons at Ohio State University. Beta and $\gamma$-ray decay curves were measured following the irradiation. ``The order of magnitudes, therefore, clearly indicate that the 23.3-hour activity is produced from Zr$^{96}$ by (p,n) reaction, since both the isobaric chains 95 and 97 have already been well-established by the study of fission fragments.'' The final quoted value of 23.35(5)~h was later independently confirmed \cite{1968Ant01} and is currently the accepted half-life. Earlier values of 4~d and 2.8~d were private communications quoted in the 1948 Table of Isotopes \cite{1948Sea01}. These values as well as a 3~d half-life reported in the Plutonium Project \cite{1951Jac01} were evidently incorrect.

\subsection{$^{97}$Nb}\vspace{0.0cm}

$^{97}$Nb was identified by Katcoff and Finkle in the 1945 paper ``Energies of Radiations of 17h Zr$^{97}$ and 75m Nb$^{97}$'' \cite{1951Kat02}. $^{239}$Pu was irradiated at the Argonne Heavy-Water Pile and $\beta$- and $\gamma$-rays were measured following chemical separation. ``The maximum $\beta$ energies of 17h Zr$^{97}$ and 75m Nb$^{97}$ as obtained by Feather analysis of the aluminum absorption curves are 2.2 and 1.4 Mev, respectively.'' The 75~min half-life agrees with the accepted value of 72.1(7)~min. Technically, the first refereed publication was by Burgus et al. in 1950 \cite{1950Bur01}, however, as participants of the Plutonium Project, they had access and gave credit to the work of Katcoff and Finkle.

\subsection{$^{98}$Nb}\vspace{0.0cm}

``Short-lived Isotopes of Nb and Zr from Fission,'' published in 1960 by Orth and Smith, reported the first observation of $^{98}$Nb \cite{1960Ort01}. Enriched $^{235}$U was irradiated with neutrons in the Los Alamos Laboratory water boiler reactor.  $^{98}$Nb was chemically separated and measured with $\beta$- and $\gamma$-scintillation spectrometers. ``The resultant decay data, plotted in [the figure], showed components of half-life 51.5$\pm$1.0 min and 23 hr. The presence of a strong photopeak at 0.780 MeV which agrees in energy with the first excited (2$^+$) state of $^{98}$Mo as determined by Coulomb excitation, suggested that the 51.5 min activity was that of $^{98}$Nb. This tentative assignment was confirmed by (n,p) reaction on $^{98}$Mo.'' This value agrees with the half-life of 51.3(4)~m for an isomeric state.

\subsection{$^{99}$Nb}\vspace{0.0cm}

$^{99}$Nb was first observed by Duffield et al. in 1950 and reported in their paper ``Radioactivities of Nb$^{99}$, Ta$^{185}$, and W$^{185}$, and the Relative ($\gamma$,\textit{n}) and ($\gamma$,\textit{p}) Cross Sections of Mo$^{100}$'' \cite{1950Duf01}. Irradiation of enriched $^{100}$Mo with 23 MeV X-rays at the University of Illinois betatron produced $^{99}$Nb in the reaction ($\gamma$,p). ``This hitherto unknown isotope has a half-life of 2.5~min. and decays by the emission of 3.2 MeV $\beta^{-}$-rays.'' The observed half-life agrees with the 2.6(2)~min half-life of an isomeric state.

\subsection{$^{100}$Nb}\vspace{0.0cm}

In the 1967 article ``Isomeres de Courte Periode du Niobium 99 et du Niobium 100,'' H\"ubenthal et al. described the first observation of $^{100}$Nb \cite{1967Hub01}. 14-15 MeV neutrons produced at the Centre d'\'Etudes Nucl\'eaires de Grenoble irradiated an enriched $^{100}$Mo target. $^{100}$Nb was identified by measuring $\beta$- and $\gamma$-ray spectra. ``$^{100}$Nb de p\'eriode 2,5~s se d\'esint\`egre par \'emission de rayonnements $\beta$ d'\'energie maximale 4$\pm$0.5 MeV et de rayonnements $\gamma$ d'\'energie 157, 360, 400, 533 et 598 keV.'' [$^{100}$Nb decays with a period of 2.5~s by $\beta$ radiation with a maximum energy of 4 MeV and $\gamma$ rays of energy 157, 360, 400, 533 and 598 keV]. This half-life is close to the presently accepted value of 2.99(11)~s for the first excited state of $^{100}$Nb. Previously reported half-lives of 3.0(3)~min \cite{1960Ort01}, 11~min \cite{1961Tak01}, and 2.8(2)~min \cite{1966Guj01} were evidently incorrect.

\subsection{$^{101}$Nb}\vspace{0.0cm}

In 1970, Eidens et al. described the first observation of $^{101}$Nb in ``On-Line Separation and Identification of Several Short-Lived Fission Products: Decay of $^{84}$Se, $^{91}$Kr, $^{97}$Y, $^{99}$Nb, $^{99}$Zr, $^{100,101}$Nb and $^{101}$Zr'' \cite{1970Eid01}. Neutrons from the J\"ulich FRJ-2 reactor irradiated a $^{235}$U target and the fission fragments were identified with a gas-filled on-line mass separator. Beta-$\gamma$- and $\gamma$-$\gamma$-coincidences were recorded. ``A 273$\pm$3 keV line and a 399$\pm$3 keV line were both assigned to $^{101}$Nb, although they were found not to be in coincidence to each other... The results of the half-lives are 7.2$\pm$0.3 sec and 6.7$\pm$0.3 sec. This gives a mean value of 7.0$\pm$0.2 sec.'' This half-life agrees with the accepted value of 7.1(3)~s. An earlier report of a 1.0(2)~min half-life \cite{1960Ort01} was evidently incorrect.

\subsection{$^{102}$Nb}\vspace{0.0cm}

$^{102}$Nb was identified for the first time by Trautmann et al. in the 1972 paper ``Identification of Short-Lived Isotopes of Zirconium, Niobium, Molybdenum, and Technetium in Fission by Rapid Solvent Extraction Techniques'' \cite{1972Tra01}. $^{235}$U and $^{239}$Pu targets were irradiated with thermal neutrons at the Mainz Triga reactor. Following chemical separation, $\gamma$-ray spectra were recorded with a Ge(Li) detector. ``2.9-sec $^{102}$Nb. The decay curve of the 296.0 keV $\gamma$-ray indicates a niobium isotope of 2.9$\pm$0.4 sec half-life.'' This half-life is close to the accepted value of 4.3(4)~s.

\subsection{$^{103,104}$Nb}\vspace{0.0cm}

In ``Low-Energy Transitions from the Deexcitation of Spontaneous Fission Fragments of $^{252}$Cf,'' Hopkins et al. reported the first observations of $^{103}$Nb and $^{104}$Nb in 1971 \cite{1971Hop01}. The spontaneous fission of $^{252}$Cf was measured by recording X- and $\gamma$-rays with a Si(Li) and a high-resolution Ge(Li) low-energy photon detector. The observed isotopes were not discussed individually in the text and the results were summarized in a table. Gamma-lines of 164.1 keV at mass 103$\pm$0 [The actual paper quotes 130$\pm$0 and we assume this to be a typographical error] and 140.9 keV at mass 104$\pm$0 were assigned to $^{103}$Nb and $^{104}$Nb, respectively. These energies agree with the current level schemes. They had already been reported in an earlier paper \cite{1970Joh01}, however, without unique mass assignment.

\subsection{$^{105}$Nb}\vspace{0.0cm}

``Odd Neutron Nuclei near A = 100: Rotational bands in $^{103}$Mo and $^{105}$Mo populated in the $\beta^-$ decays of $^{103}$Nb and $^{105}$Nb,'' by Shizuma et al., was the first article to report the observation of $^{105}$Nb in 1984 \cite{1984Shi01}. Fission products from thermal fission of $^{235}$U were studied at the J\"ulich DIDO reactor and the high-flux reactor of the ILL at Grenoble. $^{105}$Nb was identified with the fission product separators JOSEF and LOHENGRIN.  ``Least squares fits through the data for A = 105 assuming one single half-life give an average value of t$_{1/2}$ = (2.95$\pm$0.06)~s.'' This half-life is currently the sole accepted half-life for $^{105}$Nb and should be verified independently. Previously, a $\gamma$-transition of 193.6~keV had been assigned \cite{1972Hop01} incorrectly to $^{105}$Nb.

\subsection{$^{106}$Nb}\vspace{0.0cm}

$^{106}$Nb was first detected by Ahrens et al. in 1976 as described in ``Decay Properties of Neutron-Rich Niobium Isotopes'' \cite{1976Ahr02}. Thermal neutron induced fission of $^{235}$U, $^{239}$Pu, and $^{249}$Cf was investigated at the Mainz Triga reactor. $^{106}$Nb was identified by measuring $\gamma$-ray spectra with a Ge(Li) detector following chemical separation. ``With $^{239}$Pu or $^{249}$Cf as fissionable material, our niobium sample showed a weak $\gamma$-ray peak at this energy which decayed with a half-life of about 1 sec. Therefore, this half-life is assigned to $^{106}$Nb.'' This half-life agrees with the currently accepted values of 0.93(4)~s.

\subsection{$^{107}$Nb}\vspace{0.0cm}

The first observation of $^{107}$Nb was reported by \"Ayst\"o et al. in ``Discovery of Rare Neutron-Rich Zr, Nb, Mo, Tc and Ru Isotopes in Fission: Test of $\beta$ Half-life Predictions Very Far from Stability'' in 1992 \cite{1992Ays01}. At the Ion Guide Isotope Separator On-Line (IGISOL) in Jyv\"askyl\"a, Finland, targets of uranium were bombarded with 20 MeV protons. $\beta$ decays were measured with a planar Ge detector, while $\gamma$-rays were measured with a 50\% Ge detector located behind a thin plastic detector. ``The data show clearly the K$\alpha$ peaks associated with $\beta$ decay of the new isotopes $^{107}$Nb, $^{109}$Mo, $^{110}$Mo, and $^{113}$Tc.'' The reported half-life of 330(50)~ms for $^{107}$Nb agrees with the presently accepted value of 300(9)~ms.

\subsection{$^{108-110}$Nb}\vspace{0.0cm}

In 1994, Bernas et al. published the discovery of $^{108}$Nb, $^{109}$Nb, and $^{110}$Nb in ``Projectile Fission at Relativistic Velocities: A Novel and Powerful Source of Neutron-Rich Isotopes Well Suited for In-Flight Isotopic Separation'' 1994 \cite{1994Ber01}. The isotopes were produced using projectile fission of $^{238}$U at 750 MeV/nucleon on a lead target at GSI, Germany. ``Forward emitted fragments from $^{80}$Zn up to $^{155}$Ce were analyzed with the Fragment Separator (FRS) and unambiguously identified by their energy-loss and time-of-flight.'' This experiment yielded 413, 81, and 10 counts, of $^{108}$Nb, $^{109}$Nb, and $^{110}$Nb, respectively.

\subsection{$^{111-113}$Nb}\vspace{0.0cm}

$^{111}$Nb, $^{112}$Nb, and $^{113}$Nb were discovered by Bernas et al. in 1997, as reported in ``Discovery and Cross-Section Measurement of 58 New Fission Products in Projectile-Fission of 750$\cdot$AMeV $^{238}$U'' \cite{1997Ber01}. The experiment was performed using projectile fission of $^{238}$U at 750~MeV/nucleon on a beryllium target at GSI in Germany. ``Fission fragments were separated using the fragment separator FRS tuned in an achromatic mode and identified by event-by-event measurements of $\Delta$E-B$\rho$-ToF and trajectory.'' During the experiment, individual counts for $^{111}$Nb (150), $^{112}$Nb (22) and $^{113}$Nb (8) were recorded.

\subsection{$^{114,115}$Nb}\vspace{0.0cm}

The discovery of $^{114}$Nb and $^{115}$Nb was reported in the 2010 article ``Identification of 45 New Neutron-Rich Isotopes Produced by In-Flight Fission of a $^{238}$U Beam at 345 MeV/nucleon,'' by Ohnishi et al. \cite{2010Ohn01}. The experiment was performed at the RI Beam Factory at RIKEN, where the new isotopes were created by in-flight fission of a 345 MeV/nucleon $^{238}$U beam on a beryllium target. $^{114}$Nb and $^{115}$Nb were separated and identified with the BigRIPS superconducting in-flight separator. The results for the new isotopes discovered in this study were summarized in a table. Fifteen individual counts for $^{114}$Nb and four counts for $^{115}$Nb were recorded.

\section{Discovery of $^{86-120}$Tc}

The element technetium was discovered in 1937 by Perrier and Segr\`e \cite{1937Per01}. The name technetium was suggested in 1947 \cite{1947Per01} and accepted in 1949, at the 15$^{th}$ Conference of the International Union of Chemistry \cite{1949UIC01}. Until then it was referred to simply as ``element 43''. An earlier claim of discovery with the suggested name masurium \cite{1925Ber01} has been controversial for a long time. A recent historical overview of the discovery of the element technetium includes the history of the naming of technetium \cite{2005Zin01}. The first unique identification of a technetium isotope ($^{99}$Tc) was achieved in 1938 by Segr\`e and Seaborg \cite{1938Seg01}.

Thirty-five technetium isotopes from A = 86--120 have been discovered so far; these include 12 neutron-deficient and 22 neutron-rich isotopes. According to the HFB-14 model \cite{2007Gor01}, $^{140}$Tc should be the last odd-odd particle stable neutron-rich nucleus while the odd-even particle stable neutron-rich nuclei should continue through $^{137}$Tc. The proton dripline has most likely been reached at $^{86}$Tc, because $^{85}$Tc was shown to be particle unstable \cite{1999Jan01}. However, five more isotopes $^{81-85}$Tc could possibly still have half-lives longer than 10$^{-9}$~ns \cite{2004Tho01}. Thus, about 30 isotopes have yet to be discovered corresponding to 47\% of all possible technetium isotopes.

Figure \ref{f:year-nb} summarizes the year of first discovery for all technetium isotopes identified by the method of discovery. The range of isotopes predicted to exist is indicated on the right side of the figure. The radioactive technetium isotopes were produced using fusion evaporation reactions (FE), light-particle reactions (LP), neutron induced fission (NF), charged-particle induced fission (CPF), spontaneous fission (SF), and projectile fragmentation or fission (PF). Light particles also include neutrons produced by accelerators. In the following, the discovery of each technetium isotope is discussed in detail.

\begin{figure}
	\centering
	\includegraphics[scale=.5]{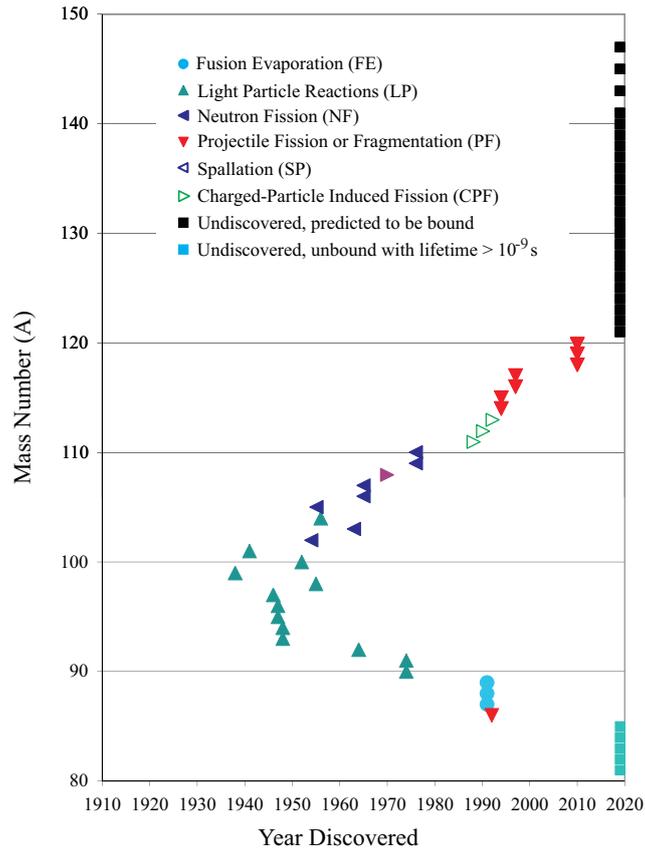}
	\caption{Technetium isotopes as a function of time when they were discovered. The different production methods are indicated. The solid black squares on the right hand side of the plot are isotopes predicted to be bound by the HFB-14 model. On the proton-rich side the light blue squares correspond to unbound isotopes predicted to have lifetimes larger than $\sim 10^{-9}$~s.}
\label{f:year}
\end{figure}

\subsection{$^{86}$Tc}\vspace{0.0cm}

The discovery of $^{86}$Tc is credited to Yennello et al. with their 1992 paper ``New Nuclei Along the Proton-Drip Line Near Z = 40,'' \cite{1992Yen01}.  At the National Superconducting Cyclotron Laboratory at Michigan State University, a 70~MeV/A $^{92}$Mo beam was produced by the K1200 cyclotron and impinged on a $^{58}$Ni target. $^{86}$Tc was identified with the A1200 fragment analyzer by measuring the time-of-flight and energy loss of the fragments. ``The mass spectra for residues with Z from 39 to 44 are shown in [the figure] with the new isotopes marked by arrows. Although both $^{84}$Mo and $^{86}$Mo have been previously observed, no reference to the identification of $^{85}$Mo was found. The other new isotopes observed in this study are $^{78}$Y, $^{82}$Nb, $^{86}$Tc, and $^{89,90}$Ru.''

\subsection{$^{87,88}$Tc}\vspace{0.0cm}

In ``New Nuclei Near the Proton Drip Line Around Z = 40,'' Rudolph et al. describes the first observation of $^{87}$Tc and $^{88}$Tc in 1991 \cite{1991Rud01}. Enriched $^{58}$Ni and $^{40}$Ca targets were bombarded with 110 MeV $^{32}$S and 170 MeV beams, respectively. $^{87}$Tc and $^{88}$Tc were formed in (p,2n) and (p,n) fusion evaporation reactions, respectively. The isotopes were separated with the Daresbury recoil separator and identified with coincident $\gamma$ rays measured with POLLYTESSA, a 19 element Compton-suppressed Ge-detector array. ``From $\gamma \gamma$ coincidence data, a decay scheme of 11 transitions was constructed for $^{88}$Tc, the ground state of which is suggested to have I$^\pi$ = 7$^-$ or 8$^+$. Two transitions identified in $^{87}$Tc follow the pattern of a g$_{9/2}$ one-particle band.''

\subsection{$^{89}$Tc}\vspace{0.0cm}

The discovery of $^{89}$Tc was published in 1991 by Heiguchi et al. in ``Half-Lives and Q$_{\beta}$ measurements for New Nuclei of $^{89}$Tc and $^{89m}$Tc'' \cite{1991Hei01}. A 95 MeV $^{32}$S beam from the Kyushu University Tandem Accelerator bombarded a nickel target and $^{89}$Tc was produced in the fusion evaporation reaction $^{60}$Ni($^{32}$S,p2n). The isotope was identified by $\beta$-$\gamma$ coincidence measurements. Further confirmation came from the reaction $^{58}$Ni($^{35}$Cl,2p2n) measured at University Tsukuba tandem-postaccelerator system. ``The half-lives of $^{89}$Tc and $^{89m}$Tc were measured to be 12.8$\pm$0.9~s and 12.9$\pm$0.8~s, respectively.'' These values are currently the only measurements and have been accepted as the half-lives for the $^{89}$Tc ground state and isomeric state. They should be verified independently.

\subsection{$^{90,91}$Tc}\vspace{0.0cm}

Iafigliola et al. reported the first observation of $^{90}$Tc and $^{91}$Tc in the 1974 paper ``The Mass Excesses of $^{91}$Tc and $^{90}$Tc'' \cite{1974Iaf01}. Targets of enriched $^{92}$Mo were bombarded with 30 and 40 MeV protons using the McGill University synchrocyclotron. $^{91}$Tc and $^{90}$Tc were identified by measuring $\beta$- and $\gamma$-rays with a plastic scintillator and a Ge(Li) detector, respectively. ``The $^{91}$Tc ground state decays with a half-life of 3.12$\pm$0.05 min, and has been assigned a J$^\pi$ of 9/2$^+$. Two isomers in $^{90}$Tc with half-lives of 7.9$\pm$0.2~s and 50$\pm$2~s have been identified for the first time.'' The half-lives for the ground states agree with the presently accepted values of 8.7(2)~s and 3.14(2)~min for $^{90}$Tc and $^{91}$Tc, respectively.

\subsection{$^{92}$Tc}\vspace{0.0cm}

The first identification of $^{92}$Tc was made in 1964 by van Lieshout et al., presented in their paper ``The Level Structure of Mo$^{92}$'' \cite{1964Van01}. An enriched $^{92}$Mo target was bombared with 17 MeV deuterons from the synchrotron at the Instituut voor Kernphysisch Onderzoek, Amsterdam. $^{92}$Tc was identified with a scintillation spectrometer following chemical separation. ``The total decay energy of Tc$^{92}$ to Mo$^{92}$ is 7.9$\pm$0.2 MeV, a value which is substantially in agreement with the one deduced from beta decay energy systematics.'' The stated half-life of 4.1(1)~min agrees with the accepted value of 4.25(15)~min. A 4.5(5)~min half-life was reported earlier and was assigned to either $^{92}$Tc or $^{93}$Tc \cite{1948Mot02}. In another measurement the 4.5~min half-life was assigned to $^{93}$Tc while a 47~min half-life was assigned to $^{92}$Tc \cite{1948Kun01}.

\subsection{$^{93}$Tc}\vspace{0.0cm}

In ``Tc$^{92}$ and Tc$^{93}$ by Relative Cross-Section Measurements,'' Kundu and Pool identified $^{93}$Tc in 1948 \cite{1948Kun01}. Natural molybdenum and enriched $^{92}$Mo, $^{94}$Mo, $^{98}$Mo targets were bombarded with 5 MeV protons and 10 MeV deuterons. $^{93}$Tc was identified by measuring charged particles, X- and $\gamma$-rays.  ``On this basis, the 47-minute activity is produced by (p,n) and the 2.7-hour activity by (p,$\gamma$) reactions, respectively. The 47-minute activity is thus  assigned to Tc$^{92}$ and the 2.7-hour activity to Tc$^{93}$.'' The reported value of 2.75(5)~h is still the  accepted half-life of $^{93}$Tc. It should be noted that the assignment of the 47~min activity to $^{92}$Tc was incorrect. The 2.7~h half-life had been measured before with no mass assignment \cite{1939Del01} or with an uncertain assignment to either $^{92}$Tc or $^{93}$Tc \cite{1948Mot02}. Also a 2~h half-life had been reported without a mass assignment \cite{1939Sea01}.

\subsection{$^{94}$Tc}\vspace{0.0cm}

In 1948, Motta and Boyd reported the first observation of $^{94}$Tc in their paper ``Characterization of Tc Activities Produced by Deuteron Bombardment of Separated Mo Isotopes'' \cite{1948Mot01}. Deuterons bombarded a $^{94}$Mo target and $^{94}$Tc was identified by measuring $\beta$- and $\gamma$ rays. ``On the basis of the yield of the activity from the molybdenum enhanced in Mo$^{94}$, the assignment of the 50-minute Tc period to mass 94 is considered probable.'' The measured half-life of 50(2)~min is included in the calculation of the accepted average value of an isomeric state. A 52-min half-life had previously been reported without a mass assignment \cite{1946Gug01,1947Gug01}.

\subsection{$^{95,96}$Tc}\vspace{0.0cm}

The discovery of $^{95}$Tc and $^{96}$Tc is credited to Edwards and Pool, for their 1947 paper ``Radioactive Isotopes of Mo and Tc'' \cite{1947Edw01}. Deuterons bombarded molybdenum targets and the isotopes were identified by $\beta$-decay curves, $\gamma$-ray spectra and X-ray photographs following chemical separation: ``Observations on the  gamma-ray decay characteristics indicate that the hard gamma-rays follow the 52-day half-life. This activity decays by K-capture with associated gamma-rays of 0.84 Mev and 0.25 Mev and is tentatively assigned to Tc$^{95}$.'' and ``The decay characteristics from Mo+p, Mo+d (Tc fraction), and Cb+$\alpha$ (Tc fraction) all show half-lives of nearly 4.3 days. This 4.3 day activity can thus be definitely assigned to Tc$^{96}$.'' These half-lives agree with the presently adopted values for an isomeric state of 61(2)~d and the ground state of 4.28(7)~d for $^{95}$Tc and $^{96}$Tc, respectively. A 62-d half-life had been reported earlier without a mass assignment \cite{1939Cac01}.

\subsection{$^{97}$Tc}\vspace{0.0cm}

Motta et al. discovered $^{97}$Tc as described in their 1946 paper ``Production and Isotopic Assignment of a 90-day activity in Element 43'' \cite{1946Mot01}. Purified samples of Ru(OH)$_{3}$ were irradiated with neutrons in the Clinton self-sustaining chain reacting pile. Beta-decays were measured following chemical separation. ``Extensive chemical tests in which the known six-hour 43$^{99}$ activity was employed as a monitor have shown this daughter activity to be an isotope of element 43, thus permitting its assignment to 43$^{97}$. The activity has been observed to decay with a half-life of 93$\pm$5 days.'' This half-life agrees with the currently accepted value of 91.0(6)~d for the isomeric state of $^{97}$Tc. Previously half-lives of 90~d \cite{1939Cac01} and 91(2)~d \cite{1941Hel02} were reported without mass assignments.

\subsection{$^{98}$Tc}\vspace{0.0cm}

In ``Production and Identification of Long-Lived Technetium Isotopes at Masses 97, 98, and 99,'' published in 1955, Boyd et al. first observed $^{98}$Tc \cite{1955Boy01}.  The Oak Ridge 86-inch cyclotron was used to bombard molydenum metal with 22 MeV protons. $^{98}$Tc was identified with a 60$^\circ$ mass spectrometer following chemical separation. ``Weighable quantities of three long-lived technetium isotopes were produced, and technetium 95, 97, and 98 isotopes have been seen on a mass spectrometer for the first time. The `missing' isotope, Tc$^{98}$, has been observed and hence must be long-lived.'' Within a month the observation of $^{98}$Tc was independently confirmed \cite{1955Kat01}.

\subsection{$^{99}$Tc}\vspace{0.0cm}

Segr\'e and Seaborg described the discovery of $^{99}$Tc in their paper ``Nuclear Isomerism in Element 43'' in 1938 \cite{1938Seg01}. Deuterons and slow neutrons were used to bombard molybdenum targets. $^{99}$Tc was identified by measuring $\beta$-, $\gamma$-, and X-rays correlating the decay to the half-life of a molybdenum isotope. ``The irradiation of molybdenum with deuterons or slow neutrons produces a radioactive molybdenum isotope with a half-life of 65 hours which emits electrons with an upper energy of approximately 1 Mev. This molybdenum decays into a second activity which has a half-life of 6 hours and which emits only a line spectrum of electrons. Since the molybdenum emits electrons, the daughter activity must be ascribed to element 43.'' Segr\'e and Seaborg do not specifically mention $^{99}$Tc; however, they quote a paper by Sagane et al., who had assigned the mentioned 65-h molybenum activity to $^{99}$Mo about two months earlier \cite{1938Sag02}.

\subsection{$^{100}$Tc}\vspace{0.0cm}

The first observation of $^{100}$Tc was reported in 1952 by Boyd et al. in their paper ``Half-Life and Radiations of Tc$^{100}$ \cite{1952Boy01}. Purified technetium metal and TcO$_2$ were irradiated with thermal neutrons in the Oak Ridge graphite pile. The resulting activities were measured with a mica end-window beta-proportional counter. ``It has been concluded that Tc$^{100}$ decays with a 15.8$\pm$0.2 second half-life, and that beta-rays of 2.8$\pm$0.2-Mev maximum energy are emitted together with gamma radiation.'' The reported half-life agrees with the presently accepted value of 15.46(19)~s.

\subsection{$^{101}$Tc}\vspace{0.0cm}

The discovery of $^{101}$Tc is credited to the 1941 paper ``Untersuchung \"uber das `19-Minuten'-Isotop von Molybd\"an und das daraus entstehende Isotop von Element 43'' by Maurer and Ramm \cite{1941Mau01}. Slow neutrons from the bombardment of beryllium with deuterons were used to activate molybdenum targets. The resulting activities were measured following chemical separation. ``Auf Grund der von uns gemessenen fast gleichen Halbwertszeiten von 14,6 Minuten f\"ur Molybd\"an und 14,0 Minuten f\"ur Element 43, lassen sich obige Widerspr\"uche als nur scheinbare jedoch leicht auffkl\"aren.'' [Based on our measurements of almost identical half-lives of 14.6~min for molybdenum and 14.0 for element 43, the above mentioned apparent contradiction can easily be resolved.] This half-life agrees with the presently accepted value of 14.22(1)~min. Hahn and Strassmann simultaneously submitted two papers reporting the same half-life for $^{101}$Tc \cite{1941Hah01,1941Hah03} giving Maurer and Ramm credit for the solution to the puzzle of the equal half-lives of $^{101}$Mo and $^{101}$Tc.

\subsection{$^{102}$Tc}\vspace{0.0cm}

Flegenheimer and Seelmann-Eggebert reported the first observation of $^{102}$Tc in the 1954 article ``\"Uber einige Isotope des Technetiums'' \cite{1954Fle01}. Chemically separated molybdenum fission fragments were used to chemically extract technetium. The remaining activities were measured with a special fast charge integrator. ``Die bisher nicht bekannte Halbwertszeit der Tochtersubstanz des unter den Kernspaltprodukten aufgefundenen Molybd\"anisotops mit einer Halbwertszeit von etwa 11 Min. (102?) wurde gemessen und betr\"agt 5$\pm$1 Sek.'' [The until now unknown half-life of the daughter substance of one of the molybdenum fission products with a half-life of about 11~min (102?) was measured and was found to be 5$\pm$1~s.] This half-life agrees with the presently accepted value of 5.28(15)~s. The mass of the mentioned molybdenum isotope had been determined less than a month earlier \cite{1954Wil01}.

\subsection{$^{103}$Tc}\vspace{0.0cm}

$^{103}$Tc was identified in 1957 by Flegenheimer and Geithoff in the paper ``\"Uber die Bestimmung des 18-and 4.5-min-Technetiums und ein neues kurzlebiges Tc-Isotop (103)'' \cite{1957Fle01}. $^{104}$Ru was irradiated with fast neutrons and the $\beta$-decay curve was measured. ``Die $\beta$-Abfallskurve zeigt nach Subtraktion des durch n,$\gamma$ gebildeten $^{105}$Ru au\ss er der 18-min-Periode noch einen kurzlebigen K\"orper von etwa 1,2 min Halbwertszeit. Dieses Nuklid war bisher unbekannt. Es handelt sich wahrscheinlich um das $^{103}$Tc, das durch den Proze\ss\  $^{104}$Ru(n,np)$^{103}$Tc entstanden ist.'' [The $\beta$-decay curve exhibits - after the subtraction of the 18-min period of $^{108}$Ru produced in the n,$\gamma$ reaction - a short-lived isotope with a half-life of about 1.2~min. This nuclide was previously unknown. It corresponds likely to $^{103}$Tc, which was produced in the process $^{104}$Ru(n,np)$^{103}$Tc.] This half-life is consistent with the presently adopted value of 54.2(8)~s. 

\subsection{$^{104}$Tc}\vspace{0.0cm}

In the 1956 paper ``Neue Spalttechnetium-Isotope (Tc-104)'', Flegenheimer and Seelmann-Eggebert correctly identified $^{104}$Tc \cite{1956Fle01}. $^{104}$Tc was produced by fast neutron irradiation of ruthenium and uranium fission induced by 28 MeV deuterons. Beta- and $\gamma$-ray spectra were recorded. ``Diese Tatsachen schlie\ss en alle Massenzahlen mit Ausnahme von 104 f\"ur das 18 min-Tc aus.'' [These facts rule out all mass numbers with the exception of 104 for the 18~min technetium half-life.] This half-life agrees with the presently adopted value of 18.3(3)~min.

\subsection{$^{105}$Tc}\vspace{0.0cm}

The first observation of $^{105}$Tc was reported by Flegenheimer et al. in the 1955 paper ``Die kurzlebigen Molybd\"an- und Technetium-Isobare der Massenzahl 105'' \cite{1955Fle01}. Ammonium diuranate was irradiated with fast neutrons and subsequently chemically separated. The $^{105}$Tc activity was indirectly determined by timed separation of $^{105}$Ru. ``Auf Grund mehrerer Versuchsreihen wurde f\"ur die Halbwertszeit des Tc-105 ein Wert von 10$\pm$1~Min. ermittelt.'' [Based on several experiments a half-life of 10$\pm$1~min was extracted for $^{105}$Tc.] This half-life is close to the currently adopted value of 7.6(1)~min.

\subsection{$^{106,107}$Tc}\vspace{0.0cm}

$^{106}$Tc and $^{107}$Tc were observed by von Baeckmann and Feuerstein as described in the 1965 paper ``\"Uber Spalt-Technetium I'' \cite{1965Von01}. $^{239}$Pu was irradiated with thermal neutrons in the Karlsruhe FR-2 reactor. The activities were identified by a $\beta$ absorption analysis following chemical separation: ``F\"ur $^{106}$Tc und $^{107}$Tc ergaben sich die Halbwertszeiten zu 37$\pm$4~s and 29$\pm$3~s.''[The extracted half-lives for $^{106}$Tc and $^{107}$Tc were 37$\pm$4~s and 29$\pm$3~s, respectively.] These half-lives are consistent with the presently accepted values of 35.6(6)~s and 21.2(2)~s.

\subsection{$^{108}$Tc}\vspace{0.0cm}

In ``A Study of the Low-Energy Transitions Arising from the Prompt De-Excitation of Fission Fragments,'' published in 1970, Watson et al. reported the first observation of $^{108}$Tc \cite{1970Wat01}. Fission fragments from the spontaneous fission of $^{252}$Cf were measured in coincidence with X-rays and conversion electrons. In a table the energy (69~keV) of the second excited state was identified correctly as it agrees with the present level scheme.

\subsection{$^{109,110}$Tc}\vspace{0.0cm}

The discovery of $^{109}$Tc and $^{110}$Tc was described by Trautmann et al. in the 1976 paper ``Identification of $^{109}$Tc and $^{110}$Tc in Fission of $^{249}$Cf'' \cite{1976Tra01}. A $^{249}$Cf target was irradiated with thermal neutrons at the Mainz Triga reactor. Gamma-ray spectra were recorded with a Ge(Li) detector following chemical separation: ``From this curve a half-life of 1.4$\pm$0.4 sec for the parent $^{109}$Tc is deduced.'' and ``The decay curve of this doublet shows two components with 5.0$\pm$0.5 and 1.0$\pm$0.2 sec half-lives. The former half-life belongs to the well known $^{108}$Tc whereas the 1.0 sec component is assigned to the decay of $^{110}$Tc.'' These half-lives are consistent with the presently accepted values of 860(40)~ms and 920(30)~ms for $^{109}$Tc and $^{110}$Tc, respectively. Earlier reported $\gamma$-ray transitions in $^{109}$Tc \cite{1970Wat01,1971Hop01} and $^{110}$Tc \cite{1972Hop01} do not agree with the present level schemes for these nuclei.

\subsection{$^{111}$Tc}\vspace{0.0cm}

Penttila et al. are credited with the first observation of $^{111}$Tc for their 1988 paper ``Half-Life Measurements for Neutron-Rich Tc, Ru, Rh, and Pd Isotopes. Identification of the New Isotopes $^{111}$Tc, $^{113}$Ru, and $^{113}$Rh'' \cite{1988Pen01}. $^{111}$Tc  was produced at the University of Jyv\"askyl\"a in Finland by proton-induced fission of $^{238}$U. The isotopes were separated with the IGISOL on-line isotope separator and identified by $\gamma$-ray, x-ray, and $\beta$-ray emissions. ``The peaks labeled as being from the previously unknown isotope $^{111}$Tc were identified through their coincidence relations with the K$\alpha$ X-ray of Ru.'' The reported half-life of 0.30(3)~s agrees with the currently accepted value of 290(20)~ms.

\subsection{$^{112}$Tc}\vspace{0.0cm}

$^{112}$Tc was discovered in 1990 by \"Ayst\"o et al. and presented in ``Collective Structure of the Neutron-Rich Nuclei, $^{110}$Ru and $^{112}$Ru'' \cite{1990Ays01}. 20~MeV protons induced fission of $^{238}$U at the IGISOL facility at the University of Jyv\"askyl\"a, Finland. $^{112}$Tc was identified by measuring $\beta$- and $\gamma$-rays of the mass separated isotopes. ``The new isotope $^{112}$Tc was found to decay with a half-life of 280(30)~ms.'' This half-life agrees with the presently adopted value of 290(20)~ms.

\subsection{$^{113}$Tc}\vspace{0.0cm}

The first observation of $^{113}$Tc was reported by \"Ayst\"o et al. in ``Discovery of Rare Neutron-Rich Zr, Nb, Mo, Tc and Ru Isotopes in Fission: Test of $\beta$ Half-life Predictions Very Far from Stability'' in 1992 \cite{1992Ays01}. At the Ion Guide Isotope Separator On-Line (IGISOL) in Jyv\"askyl\"a, Finland, targets of uranium were bombarded with 20 MeV protons. $\beta$ decays were measured with a planar Ge detector, while $\gamma$-rays were measured with a 50\% Ge detector located behind a thin plastic detector. ``The data show clearly the K$\alpha$ peaks associated with $\beta$ decay of the new isotopes $^{107}$Nb, $^{109}$Mo, $^{110}$Mo, and $^{113}$Tc.'' The reported half-life of 130(50)~ms for $^{113}$Tc agrees with the presently accepted value of 170(20)~ms.

\subsection{$^{114,115}$Tc}\vspace{0.0cm}

In 1994, Bernas et al. published the discovery of $^{114}$Tc and $^{115}$Tc in ``Projectile Fission at Relativistic Velocities: A Novel and Powerful Source of Neutron-Rich Isotopes Well Suited for In-Flight Isotopic Separation'' 1994 \cite{1994Ber01}. The isotopes were produced using projectile fission of $^{238}$U at 750 MeV/nucleon on a lead target at GSI, Germany. ``Forward emitted fragments from $^{80}$Zn up to $^{155}$Ce were analyzed with the Fragment Separator (FRS) and unambiguously identified by their energy-loss and time-of-flight.''  This experiment yielded 53 and 18 counts of $^{114}$Tc and $^{115}$Tc, respectively.

\subsection{$^{116,117}$Tc}\vspace{0.0cm}

$^{116}$Tc and $^{117}$Tc were discovered by Bernas et al. in 1997, as reported in ``Discovery and Cross-Section Measurement of 58 New Fission Products in Projectile-Fission of 750$\cdot$AMeV $^{238}$U'' \cite{1997Ber01}. The experiment was performed using projectile fission of $^{238}$U at 750~MeV/nucleon on a beryllium target at GSI in Germany. ``Fission fragments were separated using the fragment separator FRS tuned in an achromatic mode and identified by event-by-event measurements of $\Delta$E-B$\rho$-ToF and trajectory.'' During the experiment, individual counts for $^{116}$Tc (29) and $^{117}$Tc (3) were recorded.

\subsection{$^{118-120}$Tc}\vspace{0.0cm}

The discovery of $^{118}$Tc, $^{119}$Tc and $^{120}$Tc was reported in the 2010 article ``Identification of 45 New Neutron-Rich Isotopes Produced by In-Flight Fission of a $^{238}$U Beam at 345 MeV/nucleon,'' by Ohnishi et al. \cite{2010Ohn01}. The experiment was performed at the RI Beam Factory at RIKEN, where the new isotopes were created by in-flight fission of a 345 MeV/nucleon $^{238}$U beam on a beryllium target. The isotopes were separated and identified with the BigRIPS superconducting in-flight separator. The results for the new isotopes discovered in this study were summarized in a table. Twenty-seven individual counts for $^{119}$Tc and three counts for $^{120}$Tc were recorded. Additionally, clear evidence for the existence of $^{118}$Tc is shown in the A-Q spectra. The paper does not specifically claim the discovery of the isotope because of a previous publication \cite{1996Cza01}. However, this reference was only a conference proceeding and the apparent discovery of $^{118}$Tc was not included in the subsequent refereed publication \cite{1997Ber01}.

\section{Discovery of $^{87-124}$Ru}

Thirty-eight ruthenium isotopes from A = 87--124 have been discovered so far; these include 7 stable, 10 neutron-deficient and 21 neutron-rich isotopes.
According to the HFB-14 model \cite{2007Gor01}, $^{145}$Ru should be the last odd-even particle stable neutron-rich nucleus ($^{137,139,143}$Ru are calculated to be unstable) while the even-even particle stable neutron-rich nuclei should continue through $^{150}$Ru. At the proton dripline five more isotopes ($^{82-86}$Ru) are predicted to be particle stable. In addition, $^{81}$Ru could possibly still have a half-life longer than 10$^{-9}$~ns \cite{2004Tho01}. Thus, about 27 isotopes have yet to be discovered corresponding to 42\% of all possible ruthenium isotopes.

\begin{figure}
	\centering
	\includegraphics[scale=.5]{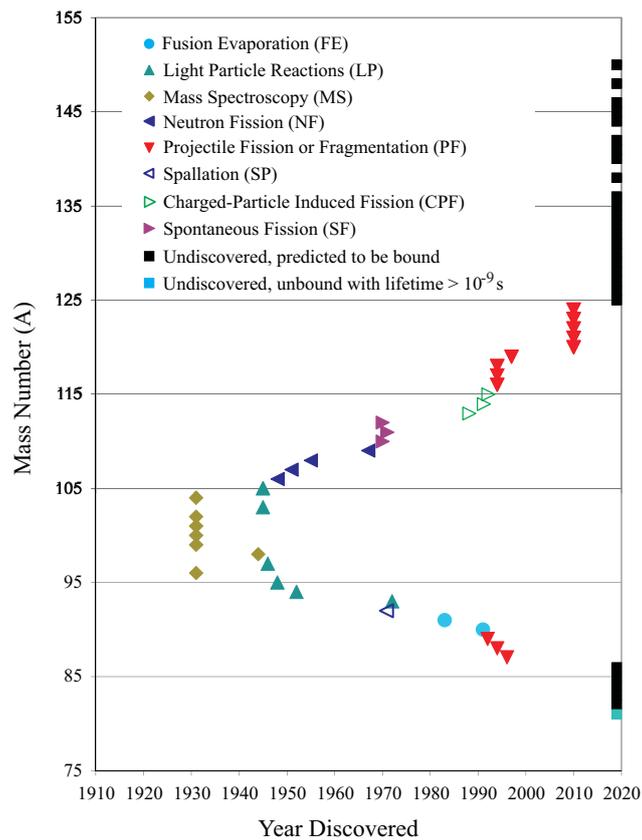}
	\caption{Ruthenium isotopes as a function of time when they were discovered. The different production methods are indicated. The solid black squares on the right hand side of the plot are isotopes predicted to be bound by the HFB-14 model. On the proton-rich side the light blue square corresponds to an unbound isotope predicted to have a lifetime larger than $\sim 10^{-9}$~s.}
\label{f:year-ru}
\end{figure}

Figure \ref{f:year-ru} summarizes the year of first discovery for all ruthenium isotopes identified by the method of discovery. The range of isotopes predicted to exist is indicated on the right side of the figure. The radioactive ruthenium isotopes were produced using fusion evaporation reactions (FE), light-particle reactions (LP), neutron induced fission (NF), charged-particle induced fission (CPF), spontaneous fission (SF), spallation reactions (SP), and projectile fragmentation or fission (PF). The stable isotopes were identified using mass spectroscopy (MS). Light particles also include neutrons produced by accelerators. In the following, the discovery of each ruthenium isotope is discussed in detail.

\subsection{$^{87}$Ru}\vspace{0.0cm}

Rykaczewski et al. discovered $^{87}$Ru in their 1995 paper ``Identification of New Nuclei at and Beyond the Proton Drip Line Near the Doubly Magic Nucleus $^{100}$Sn'' \cite{1995Ryk01}. A 63 MeV/nucleon $^{112}$Sn beam from the GANIL cyclotron complex bombarded a natural nickel target. $^{87}$Ru was identified with the Alpha and LISE3 spectrometers. ``The obtained data have allowed also for the identification of six other new nuclei, namely $^{103}$Sb, $^{104}$Sb, $^{98}$In, $^{91}$Pd, $^{89}$Rh, and $^{87}$Ru, which are clearly isolated from the neighboring heavier isotopes in the mass spectra of [the figure].'' In an earlier paper, a small number of events possibly corresponding to $^{87}$Ru were found, but the authors judged these findings as inconclusive \cite{1994Hen01}.

\subsection{$^{88}$Ru}\vspace{0.0cm}

In ``Identification of new nuclei near the proton drip line,'' Hencheck et al. reported the discovery of $^{88}$Ru in 1994 \cite{1994Hen01}. A $^{106}$Cd beam accelerated to 60 MeV/u at the National Superconducting Cyclotron Laboratory (NSCL) at Michigan State University bombarded a natural nickel target. $^{88}$Ru was analyzed with the A1900 projectile fragment separator and identified event-by-event with measurements of the magnetic rigidity, time of flight, energy-loss, and total energy. ``A number of new nuclides were identified including $^{88}$Ru, $^{90,91,92,93}$Rh, $^{92,93}$Pd, and $^{94,95}$Ag.''

\subsection{$^{89}$Ru}\vspace{0.0cm}

The discovery of $^{89}$Ru is credited to Yennello et al. with their 1992 paper ``New Nuclei Along the Proton-Drip Line Near Z = 40,'' \cite{1992Yen01}.  At the National Superconducting Cyclotron Laboratory at Michigan State University, a 70~MeV/A $^{92}$Mo beam was produced by the K1200 cyclotron and impinged on a $^{58}$Ni target.  $^{89}$Ru was identified with the A1200 fragment analyzer by measuring the time-of-flight and energy loss of the fragments. ``The mass spectra for residues with Z from 39 to 44 are shown in [the figure] with the new isotopes marked by arrows. Although both $^{84}$Mo and $^{86}$Mo have been previously observed, no reference to the identification of $^{85}$Mo was found. The other new isotopes observed in this study are $^{78}$Y, $^{82}$Nb, $^{86}$Tc, and $^{89,90}$Ru.'' The authors also claimed the discovery of $^{90}$Ru; they apparently were not aware of the previous publications by Zhou et al. \cite{1991Zho01}.

\subsection{$^{90}$Ru}\vspace{0.0cm}

In the 1991 paper ``Searching for new neutron deficient nuclide $^{90}$Ru'' Zhou et al. reported the first observation of $^{90}$Ru \cite{1991Zho01}. An enriched $^{58}$Ni target was bombarded with a 115 MeV $^{35}$Cl beam from the CIAE HI-13 tandem accelerator. X-ray and $\gamma$-rays of the residual nuclei were measured with a Si(Li) and HPGE detector, respectively. ``Thus, the 992 keV and 1002 keV $\gamma$-rays in [the figure] possibly come from the $\beta^+$ decay daughter $^{90}$Tc of a new nuclide $^{90}$Ru, which has a half-life of 13$\pm$5~s.'' This half-life is consistent with the presently adopted value of 11(3)~s.

\subsection{$^{91}$Ru}\vspace{0.0cm}

``The Decays of the T$_{z}$ = $\frac{3}{2}$ $\beta$-Delayed Proton Precursors $^{83}$Zr, $^{87}$Mo and $^{91m}$Ru,'' published in 1983 by Hagberg et al., described the first observation of $^{91}$Ru \cite{1983Hag01}. A 150 MeV $^{40}$Ca beam from the Chalk River MP tandem accelerator bombarded a self-supporting enriched $^{54}$Fe target and $^{91}$Ru was produced in the fusion evaporation reaction $^{54}$Fe($^{40}$Ca,2pn). Protons, X-rays and $\gamma$-rays were measured in coincidence in order to identify $^{91}$Ru. ``A proton precursor with a half-life of 7.6$\pm$0.8~s was assigned to $^{91}$Ru, tentatively to a 1/2$^-$ isomeric state.'' This half-life is included in the currently accepted average value for the isomeric state of $^{91}$Ru.

\subsection{$^{92}$Ru}\vspace{0.0cm}

Arlt et al. published the first observation of $^{92}$Ru in 1971 in ``The New Isotope $^{92}$Ru'' \cite{1971Arl01}. Silver chloride was bombarded with 660 MeV protons from the JINR synchrocyclotron. Gamma-ray spectra were recorded with a Ge(Li) detector following chemical separation. ``It is concluded that on the basis of the analysis of these data that the half-life for the parent $^{92}$Ru is T$_{1/2}$ = (2.5$\pm$1)~min.'' This value is consistent with the presently adopted value of 3.65(5)~min. The results were confirmed later in the same year \cite{1972Arl01}. Only a week after the second paper, de Jesus and Neirinckx independently observed $^{92}$Ru \cite{1972deJ01}.

\subsection{$^{93}$Ru}\vspace{0.0cm}

$^{93}$Ru was identified in the 1972 paper ``Decay of $^{93,93m}$Ru and Levels of $^{93}$Tc'' by Doron and Lanford \cite{1972Dor01}. Beams of 13.5 and 16.5 MeV $^3$He from the University of Rochester tandem van de Graaff accelerator bombarded an enriched $^{92}$Mo target. $^{93}$Ru was identified by measuring $\gamma$-rays with a Ge(Li) detector. ``A 680.2$\pm$0.5 keV $\gamma$-ray following a half-life of 55$\pm$10 sec was assigned to the decay of $^{93}$Ru.'' This half-life agrees with the presently accepted value of 59.7(6)~s. Previously, a 52~s half-life was measured but could not be uniquely assigned to $^{93}$Ru \cite{1955Ate02}. Also, a proposed level scheme consisting of 6 $\gamma$-rays was attributed to either $^{93}$Ru or $^{94}$Tc \cite{1971Led01}.

\subsection{$^{94}$Ru}\vspace{0.0cm}

In ``Ruthenium 94,'' published in 1952, van der Wiel and Aten Jr presented first evidence for $^{94}$Ru \cite{1952Van01}. A molybdenum target was bombarded with 52 MeV helium ions from the Instituut voor Kernphysisch Onderzoek Philips cyclotron. $^{94}$Ru is identified by measuring the $\beta$-activity following chemical separation. ``Comparison of the activity of consecutive technetium fractions obtained by milking ruthenium preparations containing Ru$^{94}$ showed the half-life of Ru$^{94}$ to be roughly 57 minutes.'' This half-life is consistent with the accepted value of 51.8(6)~min.

\subsection{$^{95}$Ru}\vspace{0.0cm}

In 1948, ``Radioactive Isotopes of Ru and Tc'' documented the first observation of $^{95}$Ru by Eggen and Pool \cite{1948Egg01}. Sheets of molybdenum metal were bombarded with 5 MeV protons and 20 MeV $\alpha$ particles. X-rays, $\gamma$-rays and $\beta$-rays were observed following chemical separation. ``Since the 1.65-hour positron activity of Ru was produced by alpha-bombardment of molybdenum and fast neutron bombardment of ruthenium but was not produced by slow neutron bombardment of ruthenium, the 1.65-hour activity may be best assigned to Ru$^{95}$.'' The measured half-life of 1.65(5)~h is in agreement with the accepted half-life of 1.643(14)~h.

\subsection{$^{96}$Ru}\vspace{0.0cm}

In 1931, Aston reported the first observation of stable $^{96}$Ru in ``Constitution of Osmium and Ruthenium'' \cite{1931Ast03}. Ruthenium tetroxide was used in the Cavendish mass spectrograph. ``Every device was tried to eliminate the mercury lines, but only on one spectrum were they so reduced that it was possible to draw conclusions that ruthenium had six isotopes with the possibility of an extremely faint seventh. The following figures, which are only rough estimates from the photometry of the faint lines, are the best available: Mass-number (Percentage abundance): 96 (5), (98) (?), 99 (12), 100 (14), 101 (22), 102 (30), 104 (17).''

\subsection{$^{97}$Ru}\vspace{0.0cm}

In 1946, Sullivan et al. published the first study of $^{97}$Ru in ``Discovery, Identification, and Characterization of 2.8d Ru$^{97}$'' \cite{1946Sul01}. Ruthenium samples were bombarded with neutrons and deuterons and $^{97}$Ru was identified by differential absorption-decay curve techniques following chemical separation. ``The average half-life value of the short-lived component was found to be 2.8$\pm$0.3 days. Using the assumption that the activation cross sections for Ru$^{96}$ and Ru$^{102}$ did not differ greatly and that the bombardment time was effectively `indefinitely short,' it was found that the observed and predicted ratios were in much closer agreement for the assignments 2.8d Ru$^{97}$ and 42d Ru$^{103}$, than for the assignments, 42d Ru$^{97}$ and 2.8d Ru$^{103}$.'' The half-life for $^{97}$Ru agrees with the presently adopted value of 2.9(1)~d.

\subsection{$^{98}$Ru}\vspace{0.0cm}

The existence of stable $^{98}$Ru was demonstrated by Ewald in the 1944 paper ``Die relativen Isotopenh\"aufigkeiten und die Atomgewichte von Kupfer und Ruthenium'' \cite{1944Ewa01}. A new photometric method was used for the mass spectroscopic determination of the isotopic abundances. ``An der Existenz des Isotopes $^{98}$Ru besteht kein Zweifel.'' [There is no doubt about the existence of the isotope $^{98}$Ru.] Aston had previously hinted at the existence of $^{98}$Ru \cite{1931Ast03}.

\subsection{$^{99-102}$Ru}\vspace{0.0cm}

In 1931, Aston reported the first observation of stable $^{99}$Ru, $^{100}$Ru, $^{101}$Ru, and $^{102}$Ru in ``Constitution of Osmium and Ruthenium'' \cite{1931Ast03}. Ruthenium tetroxide was used in the Cavendish mass spectrograph. ``Every device was tried to eliminate the mercury lines, but only on one spectrum were they so reduced that it was possible to draw conclusions that ruthenium had six isotopes with the possibility of an extremely faint seventh. The following figures, which are only rough estimates from the photometry of the faint lines, are the best available: Mass-number (Percentage abundance): 96 (5), (98) (?), 99 (12), 100 (14), 101 (22), 102 (30), 104 (17).''

\subsection{$^{103}$Ru}\vspace{0.0cm}

$^{103}$Ru was discovered by Bohr and Hole in 1945, in their paper entitled ``Radioactivity Induced by Neutrons and Deuterons in Ruthenium'' \cite{1945Boh01}. Targets of natural ruthenium metal were bombarded with 5.5 MeV deuterons, as well as fast and slow neutrons, from the cyclotron at the Stockholm Forskningsinstitutet f\"or Fysik. The activities were measured with glass Geiger-Muller counters following chemical separation. ``It then follows that the 41 d period must be assigned to $^{103}$Ru, as the substance concerned emits negative electrons.'' The reported half-life of 41(1)~d agrees with the currently adopted value of 39.26(2)~d. Previously, half-lives of 46(3)~d \cite{1936Liv01} and 45~d \cite{1942Nis01} were reported without a mass assignment. Also, a 4~h period was assigned to $^{103}$Ru and a 45~h was assigned to $^{105}$Rh \cite{1938deV01} which was evidently incorrect.

\subsection{$^{104}$Ru}\vspace{0.0cm}

In 1931, Aston reported the first observation of stable $^{104}$Ru in ``Constitution of Osmium and Ruthenium'' \cite{1931Ast03}. Ruthenium tetroxide was used in the Cavendish mass spectrograph. ``Every device was tried to eliminate the mercury lines, but only on one spectrum were they so reduced that it was possible to draw conclusions that ruthenium had six isotopes with the possibility of an extremely faint seventh. The following figures, which are only rough estimates from the photometry of the faint lines, are the best available: Mass-number (Percentage abundance): 96 (5), (98) (?), 99 (12), 100 (14), 101 (22), 102 (30), 104 (17).''

\subsection{$^{105}$Ru}\vspace{0.0cm}

$^{105}$Ru was discovered by Bohr and Hole in 1945, in their paper entitled ``Radioactivity Induced by Neutrons and Deuterons in Ruthenium'' \cite{1945Boh01}. Targets of natural ruthenium metal were bombarded with 5.5 MeV deuterons, as well as fast and slow neutrons, from the cyclotron at the Stockholm Forskningsinstitutet f\"or Fysik. The activities were measured with glass Geiger-Muller counters following chemical separation. ``We are thus forced to assign the two nuclei [4.4 h and 37 h] to $^{105}$Ru and $^{105}$Rh.'' The reported half-life of 4.4(1)~h agrees with the currently adopted value of 4.44(2)~h. Previously, the 4~h half-life had been reported without a mass assignment \cite{1936Liv01,1942Nis01}. Also, a 4~h period was assigned to $^{103}$Ru and a 20~h was assigned to $^{105}$Ru \cite{1938deV01} which was evidently incorrect.

\subsection{$^{106}$Ru}\vspace{0.0cm}

In ``Mass Spectrographic Mass Assignment of Radioactive Isotopes,'' Hayden provided the first identification of $^{106}$Ru in 1948 \cite{1948Hay01}. A 5 $\mu$C 1-year ruthenium sample produced in the Clinton pile was placed in the ion source of a spectrograph. The mass separated ions were deposited on a photographic plate and the activities measured with a Geiger counter. ``Upon development the original plate showed the normal ruthenium spectrum and also a line at mass 106, two mass numbers above the heaviest normal isotope. The transfer plate showed one line, corresponding to the line at 106. Thus the mass of the 1-year ruthenium isotope is 106.'' A 330~d activity without a mass assignment had been reported as part of the Plutonium Project \cite{1951Gle02} and a 290~d activity was not uniquely assigned to $^{106}$Ru, the mass number was listed in brackets \cite{1948Gru01}.

\subsection{$^{107}$Ru}\vspace{0.0cm}

Glendenin described the first observation of $^{107}$Ru in ``Short-Lived Ruthenium-Rhodium Decay Chains,'' which was published in 1951 as a part of the Plutonium Project Series \cite{1951Gle01}. Uranyl nitrate was irradiated with neutrons from the Clinton pile and $^{107}$Ru was identified by measuring its activity following chemical separation. ``The available mass numbers for the 4m Ru $-$ 24m Rh chain are thus limited to 107, 108, and 110 or greater... On the basis of energy a mass assignment of 107 is highly probable.'' The measured half-life agrees with the presently accepted value of 3.75(5)~min. The 4m Ru $-$ 24m Rh decay chain had been reported earlier without a mass assignment \cite{1943Bor03}.

\subsection{$^{108}$Ru}\vspace{0.0cm}

$^{108}$Ru was identified by Baro et al. in the 1955 paper ``Eine neue Isobarenreihe 108 (110)'' \cite{1955Bar01}. Uranyl was bombarded with 28 MeV deuterons, as well as fast and slow neutrons. Activities were measured with $\beta$-counters and $\gamma$-scintillation counters following chemical separation. ``Es ist daher wahrscheinlich da\ss\ die neue Isobarenreihe der Massenzahl 108 zugerechnet werden muss obwohl allerdings die Massenzahl 110 und h\"oher denkbar w\"aren.'' [It is thus probable that the new isobaric chain must be assigned to mass 108, although mass number 110 or higher could be possible.] The assigned half-life of 4~min is consistent with the currently adopted value of 4.55(5)~min.

\subsection{$^{109}$Ru}\vspace{0.0cm}

The first measurements of $^{109}$Ru were reported in the 1967 paper ``The Half-Life of Ruthenium-109,'' by Griffiths and Fritze \cite{1967Gri01}. UO$_{2}$(NO$_{3}$)$_{2}$ was irradiated with neutrons from the McMaster reactor and $^{109}$Ru was identified by measuring the activity following chemical separation. ``The resulting $^{109}$Ru decay curve is shown in [the figure]. Analysis of the data by computer (using a Gauss-Newton iterative method) gave a value for the half-life of 34.5$\pm$2.4 sec.'' This value agrees with the presently accepted value of 34.5(10)~s. The existence of $^{109}$Ru was stipulated in an earlier paper, but no half-life or other identifying observables were presented \cite{1967Fri01}.

\subsection{$^{110}$Ru}\vspace{0.0cm}

In ``A Study of the Low-Energy Transitions Arising from the Prompt De-Excitation of Fission Fragments,'' published in 1970, Watson et al. reported the first observation of $^{110}$Ru \cite{1970Wat01}. Fission fragments from the spontaneous fission of $^{252}$Cf were measured in coincidence with X-rays and conversion electrons. In a table the energy (241~keV) of the first excited state with an upper half-life limit of 0.5~ns was identified correctly.

\subsection{$^{111}$Ru}\vspace{0.0cm}

In the 1971 paper ``Low-Energy Transitions from the Deexcitation of Spontaneous Fission Fragments of $^{252}$Cf,'' Hopkins et al. reported the observation of $^{111}$Ru \cite{1971Hop01}.  Fission fragments from the spontaneous fission of $^{252}$Cf were measured in coincidence with X-rays and low-energy $\gamma$-rays. In a table the energy (103.7~keV) of the first E1 transition was identified correctly. This energy had also been reported earlier \cite{1970Joh01}, however, no specific element assignment was made. A previous 1-2~min half-life assigned to $^{111}$Ru \cite{1971Ric01} was evidently incorrect.

\subsection{$^{112}$Ru}\vspace{0.0cm}

Evidence for $^{112}$Ru was observed by Cheifetz et al. in the 1970 paper ``Experimental Information Concerning Deformation of Neutron Rich Nuclei In the A $\sim$ 100 Region'' \cite{1970Che01}. Fission fragments from the spontaneous fission of $^{252}$Cf were observed in coincidence with X- and $\gamma$-rays measured in Ge(Li) detectors. Several isotopes were identified and the observed $\gamma$-rays are listed in a table. For $^{112}$Ru levels at 236.8 (2$^+$) and 645.7 (4$^+$), and 1062.7 keV were reported.

\subsection{$^{113}$Ru}\vspace{0.0cm}

Penttila et al. reported the first observation of $^{113}$Ru in the 1988 paper ``Half-Life Measurements for Neutron-Rich Tc, Ru, Rh, and Pd Isotopes. Identification of the New Isotopes $^{111}$Tc, $^{113}$Ru, and $^{113}$Rh'' \cite{1988Pen01}. $^{113}$Ru was produced at the University of Jyv\"askyl\"a in Finland by proton-induced fission of $^{238}$U. The isotopes were separated with the IGISOL on-line isotope separator and identified by $\gamma$-ray, x-ray, and $\beta$-ray emissions. ``Several strong transitions attributed to the beta decays of $^{113}$Rh and $^{113}$Ru were observed at the A = 113 position... Hence, these results should be taken as the first observation of the decay of $^{113}$Ru.'' The reported half-life of 0.80(10)~s agrees with the currently accepted value of 800(50)~ms. An earlier report of a 3~s half-life \cite{1978Fra01} was evidently incorrect.

\subsection{$^{114}$Ru}\vspace{0.0cm}

The discovery of $^{114}$Ru is credited to Leino et al. with their 1991 paper, ``Independent and Cumulative Yields of Very Neutron-Rich Nuclei in 20 MeV p- and 18-41 MeV d-Induced Fission of $^{238}$U'' \cite{1991Lei01}. A $^{238}$U target was bombarded with 20 MeV protons from the Jyv\"askyl\"a MC-20 cyclotron. $^{114}$Ru was separated with the IGISOL isotope separator and $\gamma$- and x-rays were measured with 20 and 25$\%$ Ge detectors as well as a 1.4 cm$^{3}$ planar Ge detector. ``The half-life of $^{114}$Ru was determined from the decay of gamma activity in coincidence with Rh K$\alpha$ x-rays. The result was 0.53$\pm$0.06 s.'' The measured half-life agrees with the presently adopted value of 0.52(5)~s.

\subsection{$^{115}$Ru}\vspace{0.0cm}

The first observation of $^{115}$Ru was reported by \"Ayst\"o et al. in ``Discovery of Rare Neutron-Rich Zr, Nb, Mo, Tc and Ru Isotopes in Fission: Test of $\beta$ Half-life Predictions Very Far from Stability'' in 1992 \cite{1992Ays01}. At the Ion Guide Isotope Separator On-Line (IGISOL) in Jyv\"askyl\"a, Finland, targets of uranium were bombarded with 20 MeV protons. $\beta$ decays were measured with a planar Ge detector, while $\gamma$-rays were measured with a 50\% Ge detector located behind a thin plastic detector. ``The new isotopes $^{105}$Zr, $^{107}$Nb, $^{109}$Mo, $^{110}$Mo, $^{113}$Tc, and$^{115}$Ru were identified through ($\beta$, K$\alpha$ x-ray) coincidences and $\beta$-delayed $\gamma$-decay.'' The reported half-life of 740(80)~ms for $^{115}$Ru corresponds to the currently adopted value.

\subsection{$^{116-118}$Ru}\vspace{0.0cm}

In 1994, Bernas et al. published the discovery of $^{116}$Ru, $^{117}$Ru, and $^{118}$Ru in ``Projectile Fission at Relativistic Velocities: A Novel and Powerful Source of Neutron-Rich Isotopes Well Suited for In-Flight Isotopic Separation'' 1994 \cite{1994Ber01}. The isotopes were produced using projectile fission of $^{238}$U at 750 MeV/nucleon on a lead target at GSI, Germany. ``Forward emitted fragments from $^{80}$Zn up to $^{155}$Ce were analyzed with the Fragment Separator (FRS) and unambiguously identified by their energy-loss and time-of-flight.'' This experiment yielded 169, 30, and 3 counts of $^{116}$Ru, $^{117}$Ru, and $^{118}$Ru, respectively.

\subsection{$^{119}$Ru}\vspace{0.0cm}

$^{119}$Ru was discovered by Bernas et al. in 1997, as reported in ``Discovery and Cross-Section Measurement of 58 New Fission Products in Projectile-Fission of 750$\cdot$AMeV $^{238}$U'' \cite{1997Ber01}. The experiment was performed using projectile fission of $^{238}$U at 750~MeV/nucleon on a beryllium target at GSI in Germany. ``Fission fragments were separated using the fragment separator FRS tuned in an achromatic mode and identified by event-by-event measurements of $\Delta$E-B$\rho$-ToF and trajectory.'' During the experiment, five individual counts were recorded for $^{116}$Ru.

\subsection{$^{120-124}$Ru}\vspace{0.0cm}

The discovery of $^{120}$Ru, $^{121}$Ru, $^{122}$Ru, $^{123}$Ru, and $^{124}$Ru was reported in the 2010 article ``Identification of 45 New Neutron-Rich Isotopes Produced by In-Flight Fission of a $^{238}$U Beam at 345 MeV/nucleon,'' by Ohnishi et al. \cite{2010Ohn01}. The experiment was performed at the RI Beam Factory at RIKEN, where the new isotopes were created by in-flight fission of a 345 MeV/nucleon $^{238}$U beam on a beryllium target. The isotopes were separated and identified with the BigRIPS superconducting in-flight separator.The results for the new isotopes discovered in this study were summarized in a table. Individual counts of 143, 15, 3, and 1 for $^{121}$Ru, $^{122}$Ru, $^{123}$Ru, and $^{124}$Ru, respectively were recorded. The observation of only one $^{124}$Ru event should be considered tentative until it is confirmed by an independent measurement. The paper also shows clear evidence for the existence of $^{120}$Ru as shown in the A-Q spectra. The authors do not specifically claim the discovery of the isotope because of a previous publication \cite{1996Cza01}. However, this reference was only a conference proceeding and the apparent discovery of $^{120}$Ru was not included in the subsequent refereed publication \cite{1997Ber01}.

\section{Summary}
The discoveries of the known yttrium, zirconium, niobium, technetium, and ruthenium isotopes have been compiled and the methods of their production discussed.
The identification of these isotopes was challenging due to the presence of many isomers and especially for the isotopes near stability many quoted half-lives were incorrect or assigned to the wrong isotope. Specifically the yttrium isotopes $^{82-84}$Y,  $^{86}$Y, $^{95,96}$Y, and $^{101}$Y were initially identified incorrectly and it took 28 years before the $^{82}$Y was correctly identified. $^{88}$Y was initially assigned to $^{86}$Y and $^{93}$Y to $^{95}$Y. In addition, the half-lives of $^{87}$Y, and $^{91-94}$Y were first reported without a mass assignment.

For as many as ten zirconium isotopes ($^{81-84}$Zr, $^{93}$Zr, $^{95}$Zr, $^{99}$Zr, $^{100}$Zr, $^{101}$Zr, and $^{102}$Zr) a wrong half-life was first reported. Also, the half-lives for $^{83}$Zr, $^{98}$Zr, and $^{104}$Zr were first reported without a mass assignment.

$^{85}$Nb, $^{88}$Nb, $^{95,96}$Nb, $^{100,101}$Nb, and $^{105}$Nb were at first misidentified, and no mass assignments were possible for the first observation of the half-lives of $^{94}$Nb and $^{103,104}$Nb.

In technetium, only the half-life of $^{92}$Tc was initially misidentified and assigned to $^{93}$Tc, however, many of the half-lives ($^{92-94}$Tc, $^{96,97}$Tc, $^{101}$Tc, $^{103}$Tc, $^{109,110}$Tc) were reported before it was possible to make firm mass assignments.

Finally, the half-life of $^{105}$Ru was first incorrectly assigned to $^{103}$Ru; $^{111}$Ru, and $^{113}$Ru were misidentified; and the half-lives of $^{103}$Ru and $^{106,107}$Ru were reported before a mass assignment was possible.

\ack

This work was supported by the National Science Foundation under grants No. PHY06-06007 (NSCL) and PHY07-54541 (REU).

\bibliography{../isotope-discovery-references}

\newpage

\newpage

\TableExplanation

\bigskip
\renewcommand{\arraystretch}{1.0}

\section{Table 1.\label{tbl1te} Discovery of yttrium, zirconium, niobium, technetium, and ruthenium isotopes }
\begin{tabular*}{0.95\textwidth}{@{}@{\extracolsep{\fill}}lp{5.5in}@{}}
\multicolumn{2}{p{0.95\textwidth}}{ }\\

Isotope & Yttrium, zirconium, niobium, technetium, or ruthenium isotope \\
Author & First author of refereed publication \\
Journal & Journal of publication \\
Ref. & Reference \\
Method & Production method used in the discovery: \\

  & FE: fusion evaporation \\
  & LP: light-particle reactions (including neutrons) \\
  & MS: mass spectroscopy \\
  & NC: neutron capture reactions \\
  & PN: photo-nuclear reactions \\
  & NF: neutron induced fission \\
  & CPF: charged-particle induced fission \\
  & SF: spontaneous fission \\
  & SP: spallation \\
  & PF: projectile fragmentation of fission \\

Laboratory & Laboratory where the experiment was performed\\
Country & Country of laboratory\\
Year & Year of discovery \\
\end{tabular*}
\label{tableI}

\datatables 



\setlength{\LTleft}{0pt}
\setlength{\LTright}{0pt}


\setlength{\tabcolsep}{0.5\tabcolsep}

\renewcommand{\arraystretch}{1.0}

\footnotesize 

\begin{longtable}{@{\extracolsep\fill}llllllll@{}}
\caption{Discovery of Yttrium, Zirconium, Niobium, Technetium, and Ruthenium Isotopes. See page\ \pageref{tbl1te} for Explanation of Tables}
Isotope & Author & Journal & Ref. & Method & Laboratory & Country & Year\\
\hline\\
\endfirsthead\\
\caption[]{(continued)}
Isotope & Author & Journal & Ref. & Method & Laboratory & Country & Year\\
\hline\\
\endhead
$^{76}$Y & P. Kienle & Prog. Part. Nucl. Phys. &\cite{2001Kie01}& PF & GSI & Germany &2001 \\
$^{77}$Y & Z. Janas & Phys. Rev. Lett. &\cite{1999Jan01}& PF & GANIL & France &1999 \\
$^{78}$Y & S.J. Yennello & Phys. Rev. C &\cite{1992Yen01}& PF & Michigan State & USA &1992 \\
$^{79}$Y & H. Grawe & Z. Phys. A &\cite{1992Gra01}& SP & CERN & Switzerland &1992 \\
$^{80}$Y & C.J. Lister & Phys. Rev. C &\cite{1981Lis01}& FE & Brookhaven & USA &1981 \\
$^{81}$Y & C.J. Lister & Phys. Rev. C &\cite{1981Lis01}& FE & Brookhaven & USA &1981 \\
$^{82}$Y & C. Deprun & Z. Phys. A &\cite{1980Dep01}& FE & Orsay & France &1980 \\
$^{83}$Y & V. Maxia & J. Inorg. Nucl. Chem. &\cite{1962Max01}& FE & Berkeley & USA &1962 \\
$^{84}$Y & V. Maxia & J. Inorg. Nucl. Chem. &\cite{1962Max01}& FE & Berkeley & USA &1962 \\
$^{85}$Y & A.A. Caretto & J. Am. Chem. Soc. &\cite{1952Car01}& LP & Rochester & USA &1952 \\
$^{86}$Y & E.K. Hyde & Phys. Rev. &\cite{1951Hyd01}& LP & Berkeley & USA &1951 \\
$^{87}$Y & L.A. DuBridge & Phys. Rev. &\cite{1940DuB01}& LP & Rochester & USA &1940 \\
$^{88}$Y & R.J. Hayden & Phys. Rev. &\cite{1948Hay01}& NF & Chicago & USA &1948 \\
$^{89}$Y & F.W. Aston & Nature &\cite{1923Ast01}& MS & Cambridge & UK &1923 \\
$^{90}$Y & M.L. Pool & Phys. Rev. &\cite{1937Poo01}& LP & Michigan & USA &1937 \\
$^{91}$Y & W. Seelmann-Eggebert & Naturwiss. &\cite{1943See02}& LP & Berln& Germany &1943 \\
$^{92}$Y & R. Sagane & Phys. Rev. &\cite{1940Sag01}& LP & Tokyo & Japan &1940 \\
$^{93}$Y & S. Katcoff & Phys. Rev. &\cite{1948Kat01}& NF & Los Alamos & USA &1948 \\
$^{94}$Y & S. Katcoff & Phys. Rev. &\cite{1948Kat01}& NF & Los Alamos & USA &1948 \\
$^{95}$Y & J.D. Knight & J. Inorg. Nucl. Chem. &\cite{1959Kni01}& NF & Los Alamos & USA &1959 \\
$^{96}$Y & H. Gunther & Nucl. Phys. A &\cite{1975Gun01}& NF & Munich & Germany &1975 \\
$^{97}$Y & J. Eidens & Nucl. Phys. A &\cite{1970Eid01}& NF & Juelich & Germany &1970 \\
$^{98}$Y & J.W. Grueter& Phys. Lett. B &\cite{1970Gru01}& NF & Juelich & Germany &1970 \\
$^{99}$Y & M. Asghar & Nucl. Phys. A &\cite{1975Asg01}& NF & Grenoble & France &1975 \\
$^{100}$Y & B. Pfeiffer & J. Phys. (France) &\cite{1977Pfe01}& NF & Grenoble & France &1977 \\
$^{101}$Y & F.K. Wohn & Phys. Rev. Lett. &\cite{1983Woh01}& NF & Brookhaven & USA &1983 \\
$^{102}$Y & K. Shizuma & Phys. Rev. C &\cite{1983Shi01}& NF & Juelich & Germany &1983 \\
$^{103}$Y & M. Bernas & Phys. Lett. B &\cite{1994Ber01}& PF & GSI & Germany &1994 \\
$^{104}$Y & M. Bernas & Phys. Lett. B &\cite{1994Ber01}& PF & GSI & Germany &1994 \\
$^{105}$Y & M. Bernas & Phys. Lett. B &\cite{1994Ber01}& PF & GSI & Germany &1994 \\
$^{106}$Y & M. Bernas & Phys. Lett. B &\cite{1997Ber01}& PF & GSI & Germany &1997 \\
$^{107}$Y & M. Bernas & Phys. Lett. B &\cite{1997Ber01}& PF & GSI & Germany &1997 \\
$^{108}$Y & T. Ohnishi & J. Phys. Soc. Japan &\cite{2010Ohn01}& PF & RIKEN & Japan &2010 \\
$^{109}$Y & T. Ohnishi & J. Phys. Soc. Japan &\cite{2010Ohn01}& PF & RIKEN & Japan &2010 \\
 & & & & & & & \\
 & & & & & & & \\
$^{78}$Zr & P. Kienle & Prog. Part. Nucl. Phys. &\cite{2001Kie01}& PF & GSI & Germany &2001 \\
$^{79}$Zr & Z. Janas & Phys. Rev. Lett. &\cite{1999Jan01}& PF & GANIL & France &1999 \\
$^{80}$Zr & C.J. Lister & Phys. Rev. Lett. &\cite{1987Lis01}& FE & Daresbury & UK &1987 \\
$^{81}$Zr & W.X. Huang & Z. Phys. A &\cite{1997Hua01}& FE & Lanzhou & China &1997 \\
$^{82}$Zr & C.F. Liang & Z. Phys. A &\cite{1982Lia01}& SP & Orsay & France &1982 \\
$^{83}$Zr & M. Kaba & Radiochim. Acta &\cite{1974Kab01}& FE & Munich & Germany &1974 \\
$^{84}$Zr & G. Korschinek & Z. Phys. A &\cite{1977Kor01}& FE & Munich & Germany &1977 \\
$^{85}$Zr & F.D.S. Butement & J. Inorg. Nucl. Chem. &\cite{1963But01}& SP & Liverpool & UK &1963 \\
$^{86}$Zr & E.K. Hyde & Phys. Rev. &\cite{1951Hyd01}& LP & Berkeley & USA &1951 \\
$^{87}$Zr & B.E. Robertson & Phys. Rev. &\cite{1949Rob01}& LP & Ohio State & USA &1948 \\
$^{88}$Zr & E.K. Hyde & Phys. Rev. &\cite{1951Hyd01}& LP & Berkeley & USA &1951 \\
$^{89}$Zr & R. Sagane & Phys. Rev. &\cite{1938Sag02}& LP & Tokyo & Japan &1948 \\
$^{90}$Zr & F.W. Aston & Nature &\cite{1924Ast02}& MS & Cambridge & UK &1924 \\
$^{91}$Zr & F.W. Aston & Nature &\cite{1934Ast02}& MS & Cambridge & UK &1934 \\
$^{92}$Zr & F.W. Aston & Nature &\cite{1924Ast02}& MS & Cambridge & UK &1924 \\
$^{93}$Zr & E.P. Steinberg & Phys. Rev. &\cite{1950Ste01}& NF & Argonne & USA &1950 \\
$^{94}$Zr & F.W. Aston & Nature &\cite{1924Ast02}& MS & Cambridge & UK &1924 \\
$^{95}$Zr & W.E. Grummitt & Nature &\cite{1946Gru01}& NF & Chalk River & Canada &1946 \\
$^{96}$Zr & F.W. Aston & Nature &\cite{1934Ast02}& MS & Cambridge & UK &1934 \\
$^{97}$Zr & S. Katcoff & Nat. Nucl. Ener. Ser. &\cite{1951Kat02}& NF & Argonne & USA &1951 \\
$^{98}$Zr & K. H\"ubenthal & Compt. Rend. Acad. Sci. &\cite{1967Hub02}& LP & Grenoble & France &1967 \\
$^{99}$Zr & J. Eidens & Nucl. Phys. A &\cite{1970Eid01}& NF & Juelich & Germany &1970 \\
$^{100}$Zr & E. Cheifetz & Phys. Rev. Lett. &\cite{1970Che01}& SF & Berkeley & USA &1970 \\
$^{101}$Zr & N. Trautmann & Radiochim. Acta &\cite{1972Tra01}& NF & Mainz & Germany &1972 \\
$^{102}$Zr & R.L. Watson & Nucl. Phys. A &\cite{1970Wat01}& SF & Berkeley & USA &1970 \\
$^{103}$Zr & M. Graefenstedt & Z. Phys. A &\cite{1987Gra01}& NF & Grenoble & France &1987 \\
$^{104}$Zr & M.A.C. Hotchkins & Phys. Rev. Lett. &\cite{1990Hot01}& SF & Argonne & USA &1990 \\
$^{105}$Zr & J. Aysto& Phys. Rev. Lett. &\cite{1992Ays01}& CPF & Jyv\"askyl\"a & Finland &1992 \\
$^{106}$Zr & M. Bernas & Phys. Lett. B &\cite{1994Ber01}& PF & GSI & Germany &1994 \\
$^{107}$Zr & M. Bernas & Phys. Lett. B &\cite{1994Ber01}& PF & GSI & Germany &1994 \\
$^{108}$Zr & M. Bernas & Phys. Lett. B &\cite{1997Ber01}& PF & GSI & Germany &1997 \\
$^{109}$Zr & M. Bernas & Phys. Lett. B &\cite{1997Ber01}& PF & GSI & Germany &1997 \\
$^{110}$Zr & M. Bernas & Phys. Lett. B &\cite{1997Ber01}& PF & GSI & Germany &1997 \\
$^{111}$Zr & T. Ohnishi & J. Phys. Soc. Japan &\cite{2010Ohn01}& PF & RIKEN & Japan &2010 \\
$^{112}$Zr & T. Ohnishi & J. Phys. Soc. Japan &\cite{2010Ohn01}& PF & RIKEN & Japan &2010 \\
 & & & & & & & \\
 & & & & & & & \\
$^{82}$Nb & S.J. Yennello & Phys. Rev. C &\cite{1992Yen01}& PF & Michigan State & USA &1992 \\
$^{83}$Nb & T. Kuroyanagi & Nucl. Phys. A &\cite{1988Kur01}& FE & Kyushu & Japan &1988 \\
$^{84}$Nb & G. Korschinek & Z. Phys. A &\cite{1977Kor01}& FE & Munich & Germany &1977 \\
$^{85}$Nb & T. Kuroyanagi & Nucl. Phys. A &\cite{1988Kur01}& FE & Kyushu & Japan &1988 \\
$^{86}$Nb & I. Votsilka & Izv. Akad. Nauk SSSR, Ser. Fiz &\cite{1974Vot01}& SP & Dubna & Russia &1974 \\
$^{87}$Nb & T.A. Doron & Nucl. Phys. A &\cite{1971Dor01}& FE & Rochester & USA &1971 \\
$^{88}$Nb & R.G. Korteling & Phys. Rev. &\cite{1964Kor01}& SP & Berkeley & USA &1964 \\
$^{89}$Nb & R.M. Diamond & Phys. Rev. &\cite{1954Dia02}& LP & Harvard & USA &1954 \\
$^{90}$Nb & L. Jacobson & Nat. Nucl. Ener. Ser. &\cite{1951Jac01}& LP & Berkeley & USA &1951 \\
$^{91}$Nb & L. Jacobson & Nat. Nucl. Ener. Ser. &\cite{1951Jac01}& LP & Berkeley & USA &1951 \\
$^{92}$Nb & R. Sagane & Phys. Rev. &\cite{1938Sag03}& LP & Tokyo & Japan &1938 \\
$^{93}$Nb & F.W. Aston & Nature &\cite{1932Ast01}& MS & Cambridge & UK &1932 \\
$^{94}$Nb & R. Sagane & Phys. Rev. &\cite{1938Sag03}& NC & Tokyo & Japan &1938 \\
$^{95}$Nb & L. Jacobson & Nat. Nucl. Ener. Ser. &\cite{1951Jac01}& LP & Berkeley & USA &1951 \\
$^{96}$Nb & D. N. Kundu & Phys. Rev. &\cite{1949Kun01}& LP & Ohio State & USA &1949 \\
$^{97}$Nb & S. Katcoff & Nat. Nucl. Ener. Ser. &\cite{1951Kat02}& NF & Argonne & USA &1951 \\
$^{98}$Nb & C. J. Orth & J. Inorg. Nucl. Chem. &\cite{1960Ort01}& NF & Los Alamos & USA &1960 \\
$^{99}$Nb & R. B. Duffield & Phys. Rev. &\cite{1950Duf01}& PN & Illinois & USA &1950 \\
$^{100}$Nb & K. H\"ubenthal & Compt. Rend. Acad. Sci. &\cite{1967Hub01}& LP & Grenoble & France &1967 \\
$^{101}$Nb & J. Eidens & Nucl. Phys. A &\cite{1970Eid01}& NF & Juelich & Germany &1970 \\
$^{102}$Nb & N. Trautmann & Radiochim. Acta &\cite{1972Tra01}& NF & Mainz & Germany &1972 \\
$^{103}$Nb & F. F. Hopkins & Phys. Rev. C &\cite{1971Hop01}& SF & Austin & USA &1971 \\
$^{104}$Nb & F. F. Hopkins & Phys. Rev. C &\cite{1971Hop01}& SF & Austin & USA &1971 \\
$^{105}$Nb & K. Shizuma & Z. Phys. A &\cite{1984Shi01}& NF & Juelich/Grenoble & Germany/France &1984 \\
$^{106}$Nb & H. Ahrens & Phys. Rev. C &\cite{1976Ahr02}& NF & Mainz & Germany &1976 \\
$^{107}$Nb & J. Aysto& Phys. Rev. Lett. &\cite{1992Ays01}& CPF & Jyv\"askyl\"a & Finland &1992 \\
$^{108}$Nb & M. Bernas & Phys. Lett. B &\cite{1994Ber01}& PF & GSI & Germany &1994 \\
$^{109}$Nb & M. Bernas & Phys. Lett. B &\cite{1994Ber01}& PF & GSI & Germany &1994 \\
$^{110}$Nb & M. Bernas & Phys. Lett. B &\cite{1994Ber01}& PF & GSI & Germany &1994 \\
$^{111}$Nb & M. Bernas & Phys. Lett. B &\cite{1997Ber01}& PF & GSI & Germany &1997 \\
$^{112}$Nb & M. Bernas & Phys. Lett. B &\cite{1997Ber01}& PF & GSI & Germany &1997 \\
$^{113}$Nb & M. Bernas & Phys. Lett. B &\cite{1997Ber01}& PF & GSI & Germany &1997 \\
$^{114}$Nb & T. Ohnishi & J. Phys. Soc. Japan &\cite{2010Ohn01}& PF & RIKEN & Japan &2010 \\
$^{115}$Nb & T. Ohnishi & J. Phys. Soc. Japan &\cite{2010Ohn01}& PF & RIKEN & Japan &2010 \\
 & & & & & & & \\
 & & & & & & & \\
$^{86}$Tc & S.J. Yennello & Phys. Rev. C &\cite{1992Yen01}& PF & Michigan State & USA &1992 \\
$^{87}$Tc & D. Rudolph & J. Phys. G &\cite{1991Rud01}& FE & Daresbury & UK &1991 \\
$^{88}$Tc & D. Rudolph & J. Phys. G &\cite{1991Rud01}& FE & Daresbury & UK &1991 \\
$^{89}$Tc & K. Heiguchi & Z. Phys. A &\cite{1991Hei01}& FE & Kyushu & Japan &1991 \\
$^{90}$Tc & R. Iafigliola & Can. J. Phys. &\cite{1974Iaf01}& LP & McGill & Canada &1974 \\
$^{91}$Tc & R. Iafigliola & Can. J. Phys. &\cite{1974Iaf01}& LP & McGill & Canada &1974 \\
$^{92}$Tc & R. van Lieshout & Phys. Lett. &\cite{1964Van01}& LP & Amsterdam & Netherlands &1964 \\
$^{93}$Tc & D.N. Kundu & Phys. Rev. &\cite{1948Kun01}& LP & Ohio State & USA &1948 \\
$^{94}$Tc & E.E. Motta & Phys. Rev. &\cite{1948Mot01}& LP & Oak Ridge & USA &1948 \\
$^{95}$Tc & J.E. Edwards & Phys. Rev. &\cite{1947Edw01}& LP & Ohio State & USA &1947 \\
$^{96}$Tc & J.E. Edwards & Phys. Rev. &\cite{1947Edw01}& LP & Ohio State & USA &1947 \\
$^{97}$Tc & E.E. Motta & Phys. Rev. &\cite{1946Mot01}& LP & Oak Ridge & USA &1946 \\
$^{98}$Tc & G.E. Boyd & Phys. Rev. &\cite{1955Boy01}& LP & Oak Ridge & USA &1955 \\
$^{99}$Tc & E. Segre & Phys. Rev. &\cite{1938Seg01}& LP & Berkeley & USA &1938 \\
$^{100}$Tc & G.E. Boyd & Phys. Rev. &\cite{1952Boy01}& LP & Oak Ridge & USA &1952 \\
$^{101}$Tc & W. Maurer & Naturwiss. &\cite{1941Mau01}& LP & Berlin & Germany &1941 \\
$^{102}$Tc & J. Flegenheimer & Z. Naturforsch. &\cite{1954Fle01}& NF & Buenos Aires & Argentina &1954 \\
$^{103}$Tc & J. Flegenheimer & Z. Naturforsch. &\cite{1957Fle01}& LP & Buenos Aires & Argentina &1957 \\
$^{104}$Tc & J. Flegenheimer & Z. Naturforsch. &\cite{1956Fle01}& LP & Buenos Aires & Argentina &1956 \\
$^{105}$Tc & J. Flegenheimer & Z. Naturforsch. &\cite{1955Fle01}& NF & Buenos Aires & Argentina &1955 \\
$^{106}$Tc & A. von Beckmann & Radiochim. Acta &\cite{1965Von01}& NF & Karlsruhe & Germany &1965 \\
$^{107}$Tc & A. von Beckmann & Radiochim. Acta &\cite{1965Von01}& NF & Karlsruhe & Germany &1965 \\
$^{108}$Tc & R.L. Watson & Nucl. Phys. A &\cite{1970Wat01}& SF & Berkeley & USA &1970 \\
$^{109}$Tc & N. Trautmann & Phys. Rev. C &\cite{1976Tra01}& NF & Mainz & Germany &1976 \\
$^{110}$Tc & N. Trautmann & Phys. Rev. C &\cite{1976Tra01}& NF & Mainz & Germany &1976 \\
$^{111}$Tc & H. Penttil\"a & Phys. Rev. C &\cite{1988Pen01}& CPF & Jyv\"askyl\"a & Finland &1988 \\
$^{112}$Tc & J. Aysto& Nucl. Phys. A &\cite{1990Ays01}& CPF & Jyv\"askyl\"a & Finland &1990 \\
$^{113}$Tc & J. Aysto& Phys. Rev. Lett. &\cite{1992Ays01}& CPF & Jyv\"askyl\"a & Finland &1992 \\
$^{114}$Tc & M. Bernas & Phys. Lett. B &\cite{1994Ber01}& PF & GSI & Germany &1994 \\
$^{115}$Tc & M. Bernas & Phys. Lett. B &\cite{1994Ber01}& PF & GSI & Germany &1994 \\
$^{116}$Tc & M. Bernas & Phys. Lett. B &\cite{1997Ber01}& PF & GSI & Germany &1997 \\
$^{117}$Tc & M. Bernas & Phys. Lett. B &\cite{1997Ber01}& PF & GSI & Germany &1997 \\
$^{118}$Tc & T. Ohnishi & J. Phys. Soc. Japan &\cite{2010Ohn01}& PF & RIKEN & Japan &2010 \\
$^{119}$Tc & T. Ohnishi & J. Phys. Soc. Japan &\cite{2010Ohn01}& PF & RIKEN & Japan &2010 \\
$^{120}$Tc & T. Ohnishi & J. Phys. Soc. Japan &\cite{2010Ohn01}& PF & RIKEN & Japan &2010 \\
 & & & & & & & \\
 & & & & & & & \\
$^{87}$Ru & K. Rykaczewski & Phys. Rev. C &\cite{1995Ryk01}& PF & GANIL & France &1995 \\
$^{88}$Ru & M. Hencheck & Phys. Rev. C &\cite{1994Hen01}& PF & Michigan State & USA &1994 \\
$^{89}$Ru & S.J. Yennello & Phys. Rev. C &\cite{1992Yen01}& PF & Michigan State & USA &1992 \\
$^{90}$Ru & S. Zhou & Chin. J. Nucl. Phys. &\cite{1991Zho01}& FE & Beijing & China &1991 \\
$^{91}$Ru & E. Hagberg & Nucl. Phys. A &\cite{1983Hag01}& FE & Chalk River & Canada &1983 \\
$^{92}$Ru & R. Arlt & JETP Lett. &\cite{1971Arl01}& SP & Dubna & Russia &1971 \\
$^{93}$Ru & T.A. Doron & J. Inorg. Nucl. Chem. &\cite{1972Dor01}& LP & Rochester & USA &1972 \\
$^{94}$Ru & A. van der Wiel & Physica &\cite{1952Van01}& LP & Amsterdam & Netherlands &1952 \\
$^{95}$Ru & D.T. Eggen & Phys. Rev. &\cite{1948Egg01}& LP & Ohio State & USA &1948 \\
$^{96}$Ru & F.W. Aston & Nature &\cite{1931Ast03}& MS & Cambridge & UK &1931 \\
$^{97}$Ru & W.H. Sullivan & Phys. Rev. &\cite{1946Sul01}& LP & Chicago & USA &1946 \\
$^{98}$Ru & H. Ewald & Z.Phys. &\cite{1944Ewa01}& MS & Berlin & Germany &1944 \\
$^{99}$Ru & F.W. Aston & Nature &\cite{1931Ast03}& MS & Cambridge & UK &1931 \\
$^{100}$Ru & F.W. Aston & Nature &\cite{1931Ast03}& MS & Cambridge & UK &1931 \\
$^{101}$Ru & F.W. Aston & Nature &\cite{1931Ast03}& MS & Cambridge & UK &1931 \\
$^{102}$Ru & F.W. Aston & Nature &\cite{1931Ast03}& MS & Cambridge & UK &1931 \\
$^{103}$Ru & E. Bohr & Arkiv Mat. Astron. Fysik&\cite{1945Boh01}& LP & Stockholm & Sweden &1945 \\
$^{104}$Ru & F.W. Aston & Nature &\cite{1931Ast03}& MS & Cambridge & UK &1931 \\
$^{105}$Ru & E. Bohr & Arkiv Mat. Astron. Fysik&\cite{1945Boh01}& LP & Stockholm & Sweden &1945 \\
$^{106}$Ru & R.J. Hayden & Phys. Rev. &\cite{1948Hay01}& NF & Chicago & USA &1948 \\
$^{107}$Ru & L.E. Glendenin & Nat. Nucl. Ener. Ser. &\cite{1951Gle01}& NF & Oak Ridge & USA &1951 \\
$^{108}$Ru & G.B. Baro & Z. Naturforsch. &\cite{1955Bar01}& NF & Buenos Aires & Argentina &1955 \\
$^{109}$Ru & K. Griffith & Z. Anal. Chem. &\cite{1967Gri01}& NF & McMaster & Canada &1967 \\
$^{110}$Ru & R.L. Watson & Nucl. Phys. A &\cite{1970Wat01}& SF & Berkeley & USA &1970 \\
$^{111}$Ru & F. F. Hopkins & Phys. Rev. C &\cite{1971Hop01}& SF & Austin & USA &1971 \\
$^{112}$Ru & E. Cheifetz & Phys. Rev. Lett. &\cite{1970Che01}& SF & Berkeley & USA &1970 \\
$^{113}$Ru & H. Penttil\"a & Phys. Rev. C &\cite{1988Pen01}& CPF & Jyv\"askyl\"a & Finland &1988 \\
$^{114}$Ru & M. Leino & Phys. Rev. C &\cite{1991Lei01}& CPF & Jyv\"askyl\"a & Finland &1991 \\
$^{115}$Ru & J. Aysto& Phys. Rev. Lett. &\cite{1992Ays01}& CPF & Jyv\"askyl\"a & Finland &1992 \\
$^{116}$Ru & M. Bernas & Phys. Lett. B &\cite{1994Ber01}& PF & GSI & Germany &1994 \\
$^{117}$Ru & M. Bernas & Phys. Lett. B &\cite{1994Ber01}& PF & GSI & Germany &1994 \\
$^{118}$Ru & M. Bernas & Phys. Lett. B &\cite{1994Ber01}& PF & GSI & Germany &1994 \\
$^{119}$Ru & M. Bernas & Phys. Lett. B &\cite{1997Ber01}& PF & GSI & Germany &1997 \\
$^{120}$Ru & T. Ohnishi & J. Phys. Soc. Japan &\cite{2010Ohn01}& PF & RIKEN & Japan &2010 \\
$^{121}$Ru & T. Ohnishi & J. Phys. Soc. Japan &\cite{2010Ohn01}& PF & RIKEN & Japan &2010 \\
$^{122}$Ru & T. Ohnishi & J. Phys. Soc. Japan &\cite{2010Ohn01}& PF & RIKEN & Japan &2010 \\
$^{123}$Ru & T. Ohnishi & J. Phys. Soc. Japan &\cite{2010Ohn01}& PF & RIKEN & Japan &2010 \\
$^{124}$Ru & T. Ohnishi & J. Phys. Soc. Japan &\cite{2010Ohn01}& PF & RIKEN & Japan &2010 \\
 \\
\end{longtable}

\end{document}